\RequirePackage{fix-cm}
\documentclass[twocolumn,epjc3]{svjour3}
\smartqed  
\RequirePackage{graphicx}
\RequirePackage{amssymb}
\RequirePackage{mathtools}
\newcommand{\ket}[1]{\lvert #1 \rangle}
\newcommand{\bra}[1]{\langle #1 \rvert}
\RequirePackage[colorlinks,citecolor=blue,urlcolor=blue,linkcolor=blue]{hyperref}
\journalname{Eur. Phys. J. C}
\RequirePackage{orcidlink}
\begin{document}

\title{Entanglement and firewalls in quantum circuit model of black hole evaporation 
}
\subtitle{}


\author{Yi-Chen Lu\thanksref{addr1}\orcidlink{0009-0005-0785-1793}
        \and
        Tien Hsieh\thanksref{addr1}\orcidlink{0000-0001-7199-1241}
        \and
        Da-Shin Lee\thanksref{e1,addr1}\orcidlink{0000-0003-3187-8863} 
}

\thankstext{e1}{e-mail: dslee@gms.ndhu.edu.tw}


\institute{Department of Physics, National Dong Hwa University, Hualien, Taiwan, Republic of China \label{addr1}
}

\date{\today}

\maketitle

\begin{abstract}
We reexamine the quantum circuit model of black hole evaporation proposed in a previous work (Class. Quantum Grav. 35, 235013, 2018) \cite{tok_2018}.
This tripartite model incorporates the following systems: black hole ($\mathbf{BH}$), just radiation ($\mathbf{JR}$), and early radiation ($\mathbf{ER}$).
We apply a scrambling unitary matrix with a single parameter $\theta$ to the ground state of the qubits in infalling matter toward a black hole in order to generate initial qubit states of the black hole that are more general than those in \cite{tok_2018}.
Specifically, the scrambling unitary matrix reduces to no scrambling and maximum scrambling when $\theta=0$ and $\theta=\pi/2$, respectively.
Our aim is to explore the role of quantum monogamy in the firewall formation between the black hole and radiation.
In this model, entanglement and firewall formation depend on the black hole mass $M$ and the frequency of Hawking radiation $\omega$.
For the initial state with $\theta=\pi/2$, a firewall emerges at an earlier stage of the evolution than with $\theta=0$.
We also find that a firewall structure emerges between $\mathbf{BH}$ and $\mathbf{JR}$, and that the information is carried away by radiation for all values of $M\omega$, provided that $\theta$ lies within a certain analytically determined range.
Following unitary gate dynamics, the initial black hole qubit state can be retrieved from its imprint on the final radiation state, which was originally hidden behind the black hole's horizon.
These results may provide insight into the properties of multipartite entanglement due to the different initial states in the evolution of a quantum circuit model for black hole evaporation.
\end{abstract}

\section{Introduction}
\label{intro}
The black hole information paradox is a conflict between quantum mechanics and general relativity concerning what happens to information that falls into a black hole.
According to \cite{haw_1975,haw_1976}, a black hole evaporates through thermal radiation, which only carries information about its mass, charge, and angular momentum.
This implies that the original, detailed information is lost, violating the principle of unitarity in quantum mechanics.
In \cite{page_1993_1,page_1993_2}, Page proposed calculating the von Neumann entropy, the entanglement entropy of the black hole, and the radiation.
The entropy initially increases until it reaches its maximum value at Page time, after which it decreases back to zero as the black hole disappears.
This means that the black hole starts from a pure state and ends up with a pure state.
This is called the Page curve.
The Page curve illustrates the general characteristics of black hole evaporation and the information carried away by Hawking radiation.
Black hole complementarity is a proposed solution that suggests a black hole can be described in two consistent yet complementary ways: one from an infalling observer and the other from an external observer \cite{suss_1993}.
According to this theory, information is not lost but rather encoded in the outgoing Hawking radiation, which is accessible to a distant observer.
Meanwhile, the infalling observer sees information inside the black hole without encountering a paradox.
Since the observers cannot communicate with each other, no observer would see a contradiction.
Later, Almheiri, Marolf, Polchinski, and Sully (AMPS) in \cite{pol_2013} argued that black hole complementarity does not seem sufficient and suggested the existence of a firewall that prevents information from entering the interior of a black hole at late times due to quantum monogamy.
In quantum physics, monogamy is the property of quantum entanglement that restricts it from being shared freely within a multipartite system.
The quantum circuit model provides powerful toy frameworks for studying information flow during black hole evaporation by discrete unitary evolutions in which the entanglement structure can be explicitly explored \cite{tok_2018,luo_2017,hwang_2017,ave,osu,bro}.
The work of \cite{luo_2017} applies many-body entanglement theory to study the entanglement between different parts of Hawking radiation and black holes. A toy model that deviates from maximal bipartite entanglement is proposed.
This model illustrates that an entanglement structure can be imposed on the radiation at early and late times and moreover shows that entanglement of the radiation across the horizon is not strictly forbidden. 
Therefore, a firewall is unnecessary. According to \cite{osu}, the emitted Hawking radiation is modeled by local qubits, whereas the gravitational field of a black hole is introduced and modeled by a set of nonlocal qubits. As the qubits for the radiation propagate outward, they, together with the nonlocal qubits of the gravitational field, undergo a continuous unitary transformation.
In this scenario, information is transferred from the gravitational field of a black hole to Hawking radiation. 
This confirms that a firewall can be avoided. 
However, in \cite{tok_2018}, a quantum circuit model is introduced to describe the evolution of the evaporation process of the black hole with the horizon structure using qubits, initially in their ground state.
For relatively small values of $M \omega$, where $M$ is the black hole mass and $\omega$ is the radiation frequency, it is shown that the entanglement between the black hole and the radiation is not sustained at later steps of the evolution in the circuit model, resulting in the emergence of a firewall.
Conversely, the entanglement between the black hole and the radiation persists for larger values of $M\omega$, with no firewall emerging.
In this note, we will reexamine the circuit model in \cite{tok_2018} using more general qubit initial states to study the entanglement entropy of the black hole and mutual information between the black hole and Hawking radiation.
We will also compute the entanglement measure \cite{peres_1996,syu_2021}, the negativity \cite{vid_2002,he_2015}, from which to understand whether or not a firewall can emerge.

Quantum scrambling is the process by which local quantum information spreads out and becomes delocalized across many degrees of freedom within a complex quantum system.
This makes it difficult to retrieve the original information from a local part of the system, although the information is not destroyed.
The Hayden–Preskill protocol \cite{hay} is a quantum information thought experiment that describes how information dropped into a well-aged black hole can be recovered from its Hawking radiation.
In \cite{lan}, a quantum circuit model is proposed to examine the scrambling features of a given unitary process \cite{yos,yos_1}.
The maximally scrambling unitary process provides perfect teleportation.
Then, in \cite{kim_2023}, a partially scrambling unitary operator depending on a single parameter $\theta$ is introduced, where it reduces to no scrambling and maximum scrambling at $\theta=0$ and $\theta=\pi/2$, respectively.
It is found that the quantity of scrambling is proportional to the fidelity of teleportation.
Here, we adopt the circuit model proposed in \cite{tok_2018} with the causal and unitary gate dynamics according to \cite{bro,bro_2}.
We then apply this scrambling unitary matrix to the qubits in their ground state in the infalling matter toward the black hole.
This provides a more general initial black hole qubit state than in \cite{tok_2018}.
Our focus will be on the maximum scrambling of the initial qubit state with $\theta=\pi/2$, and we will compare the findings with those of \cite{tok_2018}. 
One of the key results is that a firewall structure emerges between the horizon pairs in this circuit model. 
Introducing scrambling initial states enlarges the range of $M\omega$ at which a firewall forms. Moreover, information can be carried away by radiation for all values of $M\omega$ provided that $\theta$ lies within a certain analytically determined range as given in (\ref{theta}). 
These results may provide insight into the properties of multipartite entanglement due to the different initial states in the evolution of a quantum circuit model for black hole evaporation.

The article is organized as follows.
Section \ref{sec2} provides a review of the quantum circuit model in \cite{tok_2018}.
Here we implement the scrambling unitary matrix with a parameter $\theta$ to the ground state of the qubits in order to generate more general initial black hole qubit states.
In Sect. \ref{sec3}, the Page curves are constructed and various entanglement entropies are computed.
In addition, the mutual information and negativity as an entanglement measure are also shown.
Conclusions are drawn in Sect. \ref{sec4}.
The wave functions at each step of the quantum circuit model for an arbitrary $\theta$ are presented in \ref{appendixA}.
\ref{appendixB} is devoted to analytical expressions for the reduced density matrices at steps 4–7.
These expressions are then used to derive analytical results for mutual information and negativity.

\section{Setup of quantum circuit model for Hawking radiation }
\label{sec2}
\begin{figure}
\centering
\includegraphics[width=8cm]{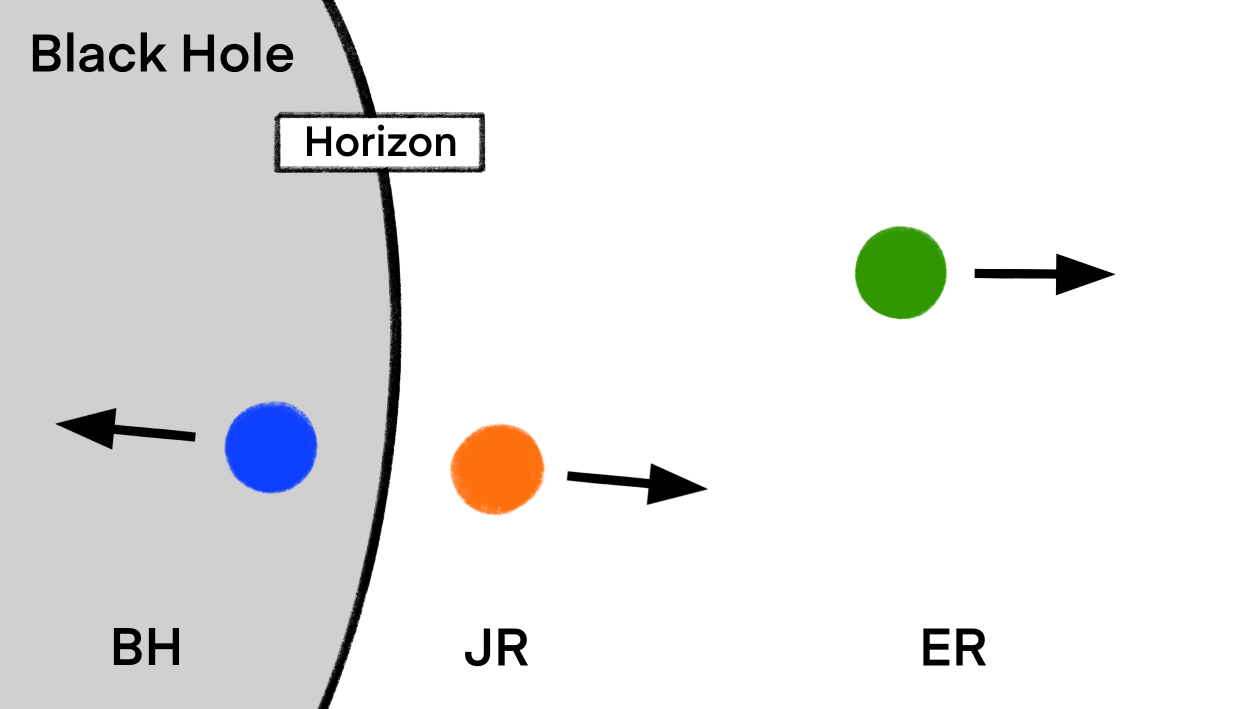}
    \caption{Schematic diagram of the qubits for the systems $\mathbf{BH}$, $\mathbf{JR}$, and $\mathbf{ER}$. }
\label{BH_evap}
\end{figure}

We start by reviewing a quantum circuit model in \cite{tok_2018}.
The model involves the degrees of freedom of the black hole ($\mathbf{BH}$), the Hawking radiation that just radiates ($\mathbf{JR}$), and the earlier Hawking radiation ($\mathbf{ER}$) using a qubit shown in Fig. \ref{BH_evap}.
We introduce a 6-qubit system for the degrees of freedom of $\mathbf{BH}$, a 1-qubit system for $\mathbf{ER}$, and a 6-qubit system for $\mathbf{ER}$.
The total Hilbert space can be specified by the product space
\begin{equation}
\mathcal{H}_{\rm tot} = \mathcal{H}_{\mathbf{BH}} \otimes \mathcal{H}_{\mathbf{JR}} \otimes \mathcal{H}_{\mathbf{ER}}.
\end{equation}
The wave function can be written as
\begin{equation}
\ket \psi = \ket {\mathbf{BH}} \otimes \ket {\mathbf{JR}} \otimes \ket {\mathbf{ER}}\, .
\end{equation}

\begin{figure}
\includegraphics[width=8cm]{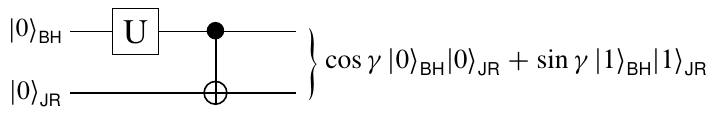}
    \caption{The CNOT-U gate }
\label{CNOT-U}
\end{figure}

The quantum state of Hawking radiation is a squeezed state, or entangled state, describing the Hawking particle and its partner pairs created at the event horizon. 
Here, in this model, the pair can comprise one qubit in $\mathbf{BH}$ and the other in $\mathbf{JR}$.
The state is often expressed as a sum over modes, representing the entanglement between the interior and exterior regions of a black hole, which for each frequency $\omega$ is given by $\sum_{n_i} e^{-4 \pi M \omega n_i} \ket {n_i}_{\mathbf{BH}} \otimes \ket {n_i}_{\mathbf{JR}}$.
Let us consider the two dominant contributions to the quantum state, namely the ground state $\ket 0$ and the one-particle state $\ket 1$.
According to \cite{tok_2018}, the squeezing parameter can be linked to the black hole mass $M$ and the Hawking radiation frequency $\omega$, which can be identified from the quantum state for Hawking radiation as
\begin{equation}
\tan \gamma= \exp ( -4 \pi M\omega) \label{gamma}
\end{equation}
where the entanglement strength is suppressed by the Boltzmann factor with the inverse of the Hawking temperature $ 8\pi M$.
Thus, the entanglement will be stronger (weaker) for small (large) values of $M \omega$.
We introduce the CNOT-U gate acting on the states of these two qubits from $\mathbf{BH}$ and $\mathbf{JR}$ to make them entangled, as shown in Fig. \ref{CNOT-U}, by introducing the matrices
\begin{align}
U &:= \begin{bmatrix}
\cos \gamma & \sin \gamma \\
\sin \gamma & -\cos \gamma
\end{bmatrix}, \quad
{\rm CNOT} := \begin{bmatrix}
\mathbb{I}_2 & 0 \\
0 & X
\end{bmatrix},\label{U}
\end{align}
where $\gamma$ is a squeezing parameter, $\mathbb{I}_n$ is an $n \times n$ identity matrix, and $X = \begin{bmatrix}
0 & 1 \\
1 & 0
\end{bmatrix}$.
The matrix representation of the CNOT-U gate is then
\begin{align}
{\rm CNOT-U} &= \begin{bmatrix}
\mathbb{I}_2 & 0 \\
0 & X
\end{bmatrix} (U \otimes \mathbb{I}_2) \notag \\
&= \begin{bmatrix}
\cos \gamma & 0 & \sin \gamma & 0 \\
0 & \cos \gamma & 0 & \sin \gamma \\
0 & \sin \gamma & 0 & -\cos \gamma\\
\sin \gamma & 0 &-\cos \gamma & 0
\end{bmatrix}\, .
\end{align}

\begin{figure}
\includegraphics[width=2.5cm]{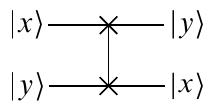}
    \caption{The SWAP gate }
\label{swap}
\end{figure}

\begin{figure*}[htp]
    \includegraphics[width=17.4cm]{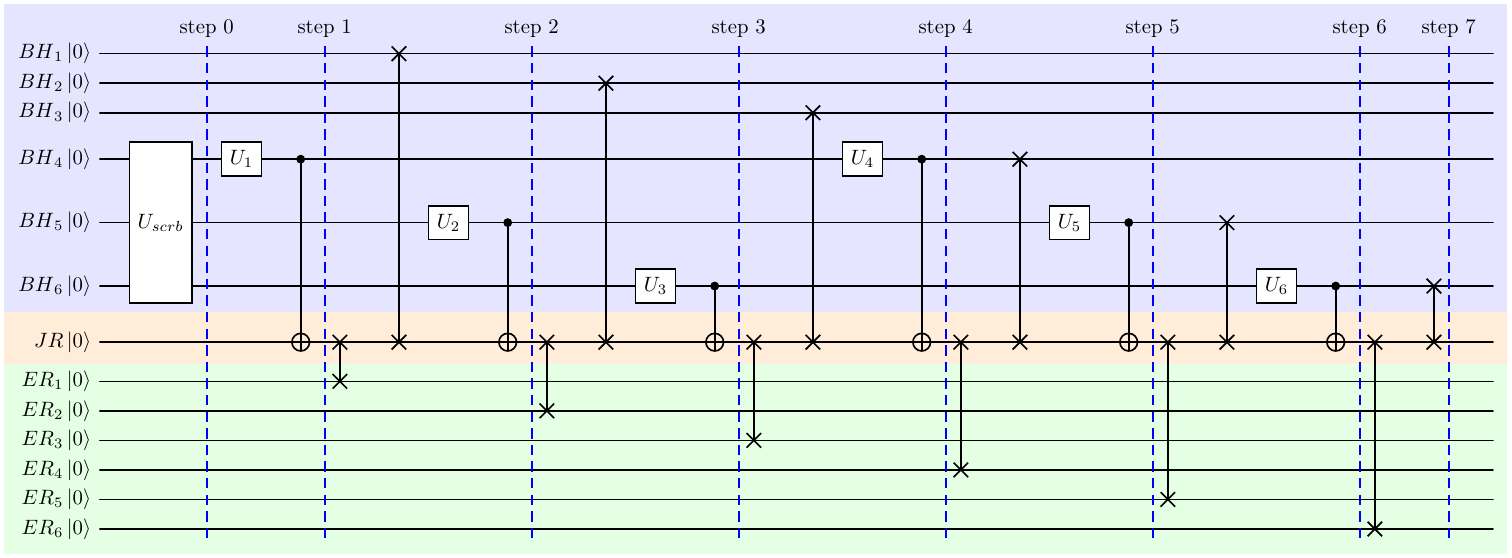}
    \caption{The Quantum circuit model with the scrambling matrix $U_{\rm scrb}$ in (\ref{ScramblingMatrix}) and the unitary matrix at step $n$, $U_n$, in (\ref{U}) }
    \label{fig:circuit}
\end{figure*}

The information flow can be mimicked using the SWAP gate shown in Fig. \ref{swap}, where the qubit in an entangled pair of $\mathbf{BH}$ and $\mathbf{JR}$ can be exchanged between $\mathbf{BH}$ ($\mathbf{JR}$) and $\mathbf{JR}$ ($\mathbf{ER}$).
In this model \cite{tok_2018} and also in \cite{bro,bro_2}, the matter collapsing into the black hole is modeled by qubits of $\mathbf{BH}$ entangled with $\mathbf{JR}$ using the CNOT-U gate, with the control $\mathbf{BH}$ and the target $\mathbf{JR}$ shown in Figs. \ref{CNOT-U} and \ref{fig:circuit}.
Later, one of the qubits falls into the black hole, carrying information with it.
The other qubit escapes from the black hole, and the information is encoded in the radiation.
This information flow can be modeled using the SWAP gate between $\mathbf{BH}$ and $\mathbf{JR}$ for the former process, and between $\mathbf{JR}$ and $\mathbf{ER}$ for the latter process, both of which are implemented in \cite{tok_2018,bro,bro_2}.

The circuit model is adopted in \cite{tok_2018} and is presented below.
In this model, the "evaporation" of a black hole does not mean that qubits disappear from the circuit.
Rather, it represents the transfer of information from the black hole subsystem and the reduction of its effective degrees of freedom.
While the total number of the qubits $ N_{\mathbf{BH}}$ remains fixed, the size of the effective Hilbert space for the black hole is reduced.
The SWAP gate between $\mathbf{JR}$ and $\mathbf{ER}$ transforms the information stored in $\mathbf{JR}$ in terms of the entangled state with $\mathbf{BH}$ to $\mathbf{ER}$, leaving $\mathbf{JR}$ in its ground state and creating an entangled state between $\mathbf{BH}$ and $\mathbf{ER}$.
The SWAP gate between $\mathbf{JR}$ and $\mathbf{BH}$ then breaks the entanglement of the qubit in $\mathbf{BH}$ with $\mathbf{ER}$, giving $\mathbf{BH}$ in the ground state and the entangled state of $\mathbf{JR}$ and $\mathbf{ER}$.
The evaporation process is described as the gradual transition of black hole qubits into the ground state with no particle excitation.
When evaporation is complete, all $\mathbf{BH}$ qubits end up in the ground state, indicating that the black hole has effectively disappeared, while the information it originally carried has been transferred to the radiation qubits.
Based on the entropy $ S \propto {M^2}$, given by the logarithm of the effective size of the Hilbert space of $\mathbf{BH}$, which is proportional to $\log_2 2^{N_{\mathbf{BH}}-(n-1)}$, for each step $n$, the mass of the black hole becomes \cite{tok_2018}
\begin{equation}
M_n=M_{\rm in} \sqrt{ 1-\frac{n-1}{N_{\mathbf{BH}}}}\,, \label{Mn}
\end{equation}
where $M_{\rm in}$ is an initial black hole mass and $N_{\mathbf{BH}}$ is the number of qubits of the black hole.
The squeezing parameter $\gamma_n$ at each step in (\ref{gamma}) can be determined using the formula of $M_n$ above.

In \cite{tok_2018}, the initial state of the qubits is chosen to be their ground state,
\begin{equation}
\ket{\rm in} = \ket{000000}_{\mathbf{BH}} \otimes \ket{0}_{\mathbf{JR}} \otimes \ket{000000}_{\mathbf{ER}} .
\end{equation}
Here, we generalize the above initial state by applying the scrambling matrix $U_{\rm scrb}$ to qubits 4–6 and keeping qubits 1–3 in the ground state in the infalling matter becoming an initial black hole qubit state.
The introduction of the $U_{\rm scrb}$ gate is designed to study the scrambling and quantum teleportation in \cite{kim_2023}.
The general form of this scrambling unitary matrix with a phase $\theta$ is given by \cite{kim_2023}
\begin{equation} \label{ScramblingMatrix}
\resizebox{8.5cm}{!}{$
U_{\rm scrb} = \frac{1}{4} \left[
\begin{array}{cccccccc}
 u_{2,+} & 0 & 0 & -u_{1,+} & 0 & -u_{1,+} & -u_{1,+} & 0 \\
 0 & u_{3,+} & -u_{1,+} & 0 & -u_{1,+} & 0 & 0 & u_{1,+} \\
 0 & -u_{1,+} & u_{3,+} & 0 & -u_{1,+} & 0 & 0 & u_{1,+} \\
 u_{1,-} & 0 & 0 & u_{3,-} & 0 & -u_{1,-} & -u_{1,-} & 0 \\
 0 & -u_{1,+} & -u_{1,+} & 0 & u_{3,+} & 0 & 0 & u_{1,+} \\
 u_{1,-} & 0 & 0 & -u_{1,-} & 0 & u_{3,-} & -u_{1,-} & 0 \\
 u_{1,-} & 0 & 0 & -u_{1,-} & 0 & -u_{1,-} & u_{3,-} & 0 \\
 0 & -u_{1,-} & -u_{1,-} & 0 & -u_{1,-} & 0 & 0 & u_{2,-} \\
\end{array}
\right]
$} \end{equation}
acting on the qubits $\mathbf{BH}_4$, $\mathbf{BH}_5$, and $\mathbf{BH}_6$, where
\begin{equation} \resizebox{8.5cm}{!}{$
u_{1,\pm} = 1 - e^{\pm 2i\theta}, \quad
u_{2,\pm} = 1 + 3e^{\pm 2i\theta}, \quad
u_{3,\pm} = 3 + e^{\pm 2i\theta}. \label{u_ipm}
$} \end{equation}
Specifically, when $\theta=0$ and $\theta=\pi/2$, there is no scrambling and maximal scrambling, respectively.
The study of the multipartite entanglement among $\mathbf{BH}$, $\mathbf{JR}$, and $\mathbf{ER}$, and the formation of the firewall between $\mathbf{BH}$ and $\mathbf{JR}$ will be reexamined by computing the various entanglement entropies ($S_\mathbf{R}$), the mutual information ($I$), and the entanglement measure, the negativity ($N$). 
Note that the scrambling is only applied to qubits 4–6, and the conclusions regarding the firewall are specific to these particular initial states. 
Further investigation is needed to understand whether scrambling quantum states could generally enhance firewall formation.

\section{Page curves and entanglement measures}
\label{sec3}
Consider the tripartite system consisting of $\mathbf{BH}$, $\mathbf{JR}$, and $\mathbf{ER}$.
The entanglement entropy of $\mathbf{A}$ can be obtained from its reduced density matrix $\rho_\mathbf{A}$ by tracing out the degrees of freedom of $\mathbf{R} \coloneqq \mathbf{B} \cup \mathbf{C}$, where $\mathbf{A}$, $\mathbf{B}$, and $\mathbf{C}$ can be $\mathbf{BH}$, $\mathbf{JR}$, or $\mathbf{ER}$.
It is given by
\begin{align}
    S_\mathbf{A} &= -\mathrm{Tr}(\rho_\mathbf{A} \log_2 \rho_\mathbf{A})
    = -\sum_{i} \lambda_{\mathbf{A}i} \log_2 \lambda_{\mathbf{A}i} \, ,
\label{S}
\end{align}
where $\lambda_{\mathbf{A}i}$ is an eigenvalue of the reduced density matrix $\rho_\mathbf{A}$.
Then one can compute the mutual information of the bipartite system defined as
\begin{align}
    I(\mathbf{A}:\mathbf{B})
    &= S_\mathbf{A} + S_\mathbf{B} - S_{\mathbf{A} \cup \mathbf{B}} \notag\\
    &=-\sum_{i} \lambda_{\mathbf{A}i} \log_2 \lambda_{\mathbf{A}i}
    -\sum_{i} \lambda_{\mathbf{B}i} \log_2 \lambda_{\mathbf{B}i} \notag\\
    &\quad +\sum_{i} \lambda_{\mathbf{A} \cup \mathbf{B}i} \log_2 \lambda_{\mathbf{A} \cup \mathbf{B}i} .
\label{I}
\end{align}
In particular, for the combined system of $\mathbf{A}$, $\mathbf{B}$, and $\mathbf{C}$ in the pure state, $S_{\mathbf{A}\cup \mathbf{B}} = S_\mathbf{C}$.
The negativity in \cite{vid_2002,he_2015} is defined from the partial transpose of the density matrix denoted by $\rho^T$.
According to \cite{peres_1996,hor_1996}, for the density matrix $\rho_{\mathbf{A} \mathbf{B}} =\sum p^{ab}_{a'b'} \ket{a} \bra{a'} \otimes \ket{b} \bra{b'}$, its partial transpose in $\mathbf{A}$ is $\rho^{T_\mathbf{A}}_{\mathbf{A} \mathbf{B}} =\sum p^{ab}_{a'b'} \ket{a'} \bra{a} \otimes \ket{b} \bra{b'}$.
For the density matrix with all positive eigenvalues, its transposed density matrix might have negative eigenvalues.
Thus, the negativity can be computed from
\begin{align}
N(\mathbf{A}:\mathbf{B}) & = \frac{1}{2} \left( \sum_i \left| \lambda_{\mathbf{A} \mathbf{B}i}^T \right| -1 \right) \,, \qquad \sum_i \lambda_{\mathbf{A} \mathbf{B}i}^T=1 \, ,\label{N}
\end{align}
where $\lambda_{\mathbf{A} \mathbf{B}i}^T$ is the eigenvalue of the density matrix $\rho^{T_\mathbf{A}}_{\mathbf{A,B}}$ or $\rho^{T_\mathbf{B}}_{\mathbf{A,B}}$.
Note that for any separable or disentangled state, all eigenvalues are positive, leading to $N=0$.
$N>0$ is a measure of how much the transposed density matrix fails to be positive definite values, which gives the strength of the entanglement of the states.
Note that positive negativity is both a necessary and sufficient condition for entanglement (i.e., non-separability) in bipartite systems such as the $\mathbf{BH}$–$\mathbf{JR}$ subsystem in this model \cite{peres_1996,hor_1996}. 
It is worth emphasizing that vanishing negativity signals the disentanglement of the horizon pairs within the circuit. However, determining whether a firewall exists to prevent information from entering a black hole for an infalling observer would require a more complete gravitational description. 
We will compute these quantities and present numerical values and analytic expressions below.
The wave function at each step is listed in \ref{appendixA}, with which the relevant reduced density with the degrees of freedom of interest can be computed by tracing out the irrelevant degrees of freedom.

\subsection{Entanglement entropy}
\begin{figure*}[htp]
\centering
\includegraphics[width=11cm]{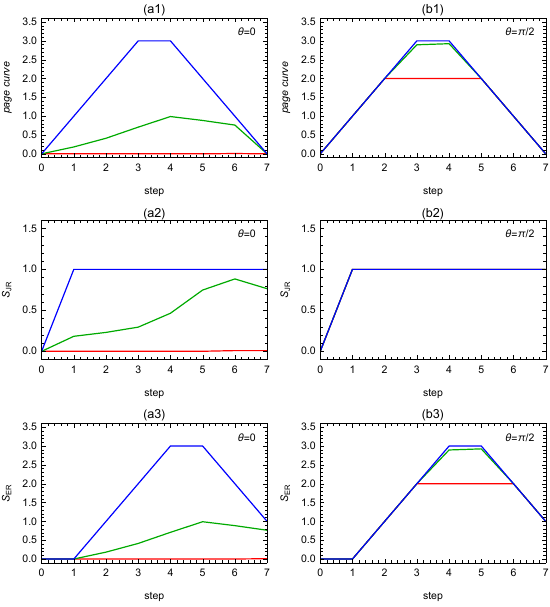}
    \caption{
    Evolution of the entanglement entropy for $\mathbf{BH}$, $\mathbf{JR}$, and $\mathbf{ER}$ at each step.
    The plots in (a) and (b) are for $\theta=0$ and $\theta=\pi/2$, respectively.
    The initial black hole masses $M_{\rm in} \omega$ are chosen to be $0.5 \sqrt{2}$ (red), $0.1 \sqrt{2}$ (green) and $0.0001 \sqrt{2}$ (blue). 
    Note that smaller values of $M \omega$ correspond to stronger entanglement between the horizon pairs }
    \label{figS}
\end{figure*}

The primary focus in this note is on $\theta=\pi/2$ in the initial state.
Here we present the entanglement entropy $S_{\mathbf{BH}}$, $S_{\mathbf{JR}}$, and $S_{\mathbf{ER}}$ using (\ref{S}).
In Fig. \ref{figS}(a), the expected Page curves of $S_{\mathbf{BH}}$ are shown with various values of the initial mass $M_{\rm in} \omega$ ($M_{\rm in} \omega=0.5\sqrt{2}$, $M_{\rm in} \omega=0.1\sqrt{2}$, and $M_{\rm in} \omega=0.0001\sqrt{2}$) for $\theta=0$ and $\pi/2$, where the curves reach their maximum at the Page time about or after step 4.
For a smaller value of $M_{\rm in} \omega$, giving rise to strong entanglement, $S_{\mathbf{BH}}$ has a higher value and has a symmetric shape resulting from the randomness of the radiation \cite{tok_2018}.
Note that in Fig. \ref{figS}(a) for $M_{\rm in} \omega=0.5\sqrt{2}$ and $\theta=0$, the value of $S_{\mathbf{BH}}$ remains small for steps 1–7, which is due to insufficient entanglement between $\mathbf{BH}$ ($\mathbf{JR}$) and $\mathbf{JR}$ ($\mathbf{ER}$) given by (\ref{gamma}).

\subsection{Mutual information and negativity}
\begin{figure*}[htp]
\centering
    \includegraphics[width=11cm]{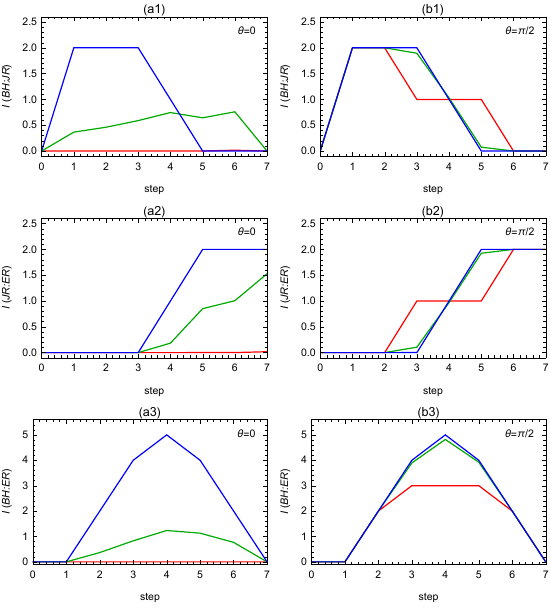}
    \caption{
    Evolution of the mutual information for $I(\mathbf{BH}:\mathbf{JR})$, $I(\mathbf{JR}:\mathbf{ER})$, and $I(\mathbf{BH}:\mathbf{ER})$ at each step.
    The plots in (a) and (b) are for $\theta=0$ and $\theta=\pi/2$, respectively.
    The initial black hole masses $M_{\rm in} \omega$ are chosen to be $0.5 \sqrt{2}$ (red), $0.1 \sqrt{2}$ (green), and $0.0001 \sqrt{2}$ (blue).
    Note that smaller values of $M \omega$ correspond to stronger entanglement between the horizon pairs }
    \label{figI}
\end{figure*}

\begin{figure*}[htp]
\centering
    \includegraphics[width=11cm]{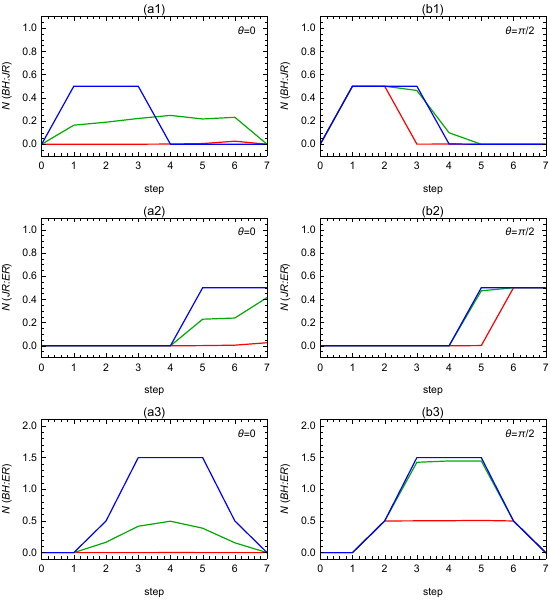}
    \caption{
    Evolution of the negativity for $N(\mathbf{BH}:\mathbf{JR})$, $N(\mathbf{JR}:\mathbf{ER})$, and $N (\mathbf{BH}:\mathbf{ER})$ at each step.
    The plots in (a) and (b) are for $\theta=0$ and $\theta=\pi/2$, respectively.
    The initial black hole masses $M_{\rm in} \omega$ are chosen to be $0.5 \sqrt{2}$ (red), $0.1 \sqrt{2}$ (green), and $0.0001 \sqrt{2}$ (blue). 
    Note that smaller values of $M \omega$ correspond to stronger entanglement between the horizon pairs }
    \label{figN}
\end{figure*}

The plots of $I(\mathbf{BH}:\mathbf{JR})$, $I(\mathbf{JR}:\mathbf{ER})$, and $I(\mathbf{BH}:\mathbf{ER})$ in (\ref{I}) are shown in Fig. \ref{figI}.
For the initial state with $\theta=\pi/2$, during the early steps, $I(\mathbf{BH}:\mathbf{JR})$ and $I(\mathbf{BH}:\mathbf{ER})$ increase from zero, reaching their respective maximum values at the Page time.
Finally, they start to decrease to zero, at which step $I(\mathbf{JR}:\mathbf{ER})$ begins to increase to a finite value.
The information is carried away by radiation, resulting in the emergence of the firewall structure between $\mathbf{BH}$ and $\mathbf{JR}$, as given by the negativity calculations for all values of $M \omega$.
In comparison, for the case of no scrambling ($\theta=0$), when $M\omega \gg 0.1$, the mutual information $I$ follows the same behavior as in $\theta=\pi/2$.
However, when $M\omega \gtrsim 1$, the value of $I(\mathbf{BH}:\mathbf{JR})$ in Fig. \ref{figI}(a) starts from zero, gradually increases to a finite value, and finally reaches zero at final step 7 when $\mathbf{BH}$ evaporates completely.
In this case, the information between $\mathbf{BH}$ and $\mathbf{JR}$ is never lost, and there is no firewall emergence to be seen from the negativity.
Again, when $M\omega \gg 0.1$ and $\theta=0$, the values of the mutual information remain small during steps 1–7.
We also present the plots of $N(\mathbf{BH}:\mathbf{JR})$, $N(\mathbf{JR}:\mathbf{ER})$, and $N(\mathbf{BH}:\mathbf{ER})$ in (\ref{N}) in Fig. \ref{figN}.
In particular, it reveals that when the mutual information between $\mathbf{BH}$ and $\mathbf{JR}$ from $I(\mathbf{BH}:\mathbf{JR})$ is (not) lost , the firewall given by $N(\mathbf{BH}:\mathbf{JR})$ does (not) form, indicating that all information is (not) carried away by radiation.
For $\theta=0$, and in the case of the relatively weak entanglement between $\mathbf{BH}$ and $\mathbf{JR}$ with $M \omega \ge 1$, the quantum monogamy places less restriction to share the entanglement with $\mathbf{ER}$, where the entanglement between any two of them can still exist.
As for the $\theta=\pi/2$ case of maximum scrambling, even in the case of $M \omega \gg 0.1$ resulting in relatively weak entanglement, quantum monogamy seems to force the breaking of the entanglement between $\mathbf{BH}$ and $\mathbf{JR}$, and the formation of a firewall between them.
Later we will show the range of $\theta$ as a function of $M\omega$ in (\ref{theta}), where the entanglement $N(\mathbf{BH}:\mathbf{JR})$ between $\mathbf{BH}$ and $\mathbf{JR}$ is not totally lost.

We now present an analysis of the mutual information and negativity during steps 4–7.
The results of $\theta=\pi/2$ will be compared with those of $\theta=0$ in \cite{tok_2018}.
The reduced densities at steps 4–7 for an arbitrary $\theta$ are presented in \ref{appendixB}. \\

\begin{figure*}[htp]
\centering
    \includegraphics[width=17.4cm]{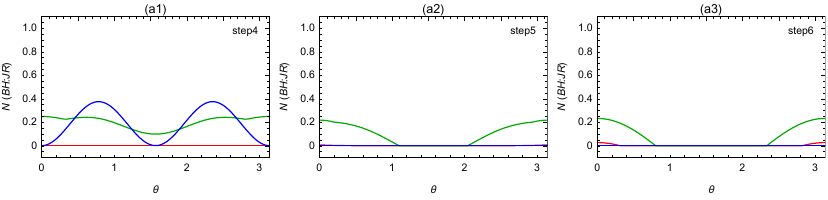}
    \caption{
    Variation of $N(\mathbf{BH}:\mathbf{JR})$ as a function of $\theta$ at each step.
    The initial black hole masses $M_{\rm in} \omega$ are chosen to be $0.5 \sqrt{2}$ (red), $0.1 \sqrt{2}$ (green), and $0.0001 \sqrt{2}$ (blue). 
    Note that smaller values of $M \omega$ correspond to stronger entanglement between the horizon pairs }
    \label{figNtheta}
\end{figure*}

\subsubsection{At step 4 for $\theta=\pi/2$}
At step 4, the wave function for $\theta=\pi/2$ is given by
\begin{equation}
\begin{split}
\ket{\psi^4}
&{=} \frac{1}{4} \Bigg\{ \Big[
\cos(\gamma_1{+}\gamma_2{+}\gamma_3)
\big( \cos{\gamma_4}\ket{000} {+} \sin{\gamma_4}\ket{110} \big) \\
&\quad+
\sin(\gamma_1+\gamma_2+\gamma_3) \\
&\qquad \big( \sin {\gamma_4}\ket{001} - \cos {\gamma_4}\ket{111} \big) \Big]_{\mathbf{BH_4},\mathbf{JR},\mathbf{ER}_{1}} \\ 
&\quad \times \big( -\ket{00}\ket{00} + \ket{11}\ket{11} \big)_{{\mathbf{BH_5},\mathbf{ER}_{2}};{\mathbf{BH_6},\mathbf{ER}_{3}}} \\ &\quad
{+} \Big[
{-}\sin(\gamma_1{+}\gamma_2{+}\gamma_3)
\big( \cos{\gamma_4}\ket{000} + \sin{\gamma_4}\ket{110} \big) \\
&\quad+ \cos(\gamma_1+\gamma_2+\gamma_3) \\
&\qquad \big( \sin {\gamma_4}\ket{001} - \cos {\gamma_4}\ket{111} \big) \Big]_{\mathbf{BH_4},\mathbf{JR},\mathbf{ER}_{1}} \\ &
\quad \times \big( \ket{00}\ket{11} + \ket{11}\ket{00} \big)_{{\mathbf{BH_5},\mathbf{ER}_{2}};{\mathbf{BH_6},\mathbf{ER}_{3}}}
\Bigg\} \\
&\quad \times \ket{0}_{\mathbf{BH_1},\mathbf{BH_2},\mathbf{BH_3},\mathbf{ER}_{4},\mathbf{ER}_{5},\mathbf{ER}_{6}} ,
\end{split}
\end{equation}
%
which is different from the wave function at
$\theta=0$ shown to be
\begin{align}
\ket{\psi^4}
=&\, \Big( \cos {\gamma_1}\cos {\gamma_4}\ket{000} + \cos {\gamma_1}\sin {\gamma_4}\ket{110} \notag \\
&\quad+\sin {\gamma_1}\sin {\gamma_4}\ket{001} \notag \\
&\qquad -\sin {\gamma_1}\cos {\gamma_4}\ket{111} \Big)_{\mathbf{BH_4},\mathbf{JR},\mathbf{ER}_{1}} \notag \\
&\quad \times \Big( \cos {\gamma_2}\ket{00} + \sin {\gamma_2}\ket{11} \Big)_{\mathbf{BH_5},\mathbf{ER}_{2}} \notag \\
&\quad \times \Big( \cos {\gamma_3}\ket{00} + \sin {\gamma_3}\ket{11} \Big)_{\mathbf{BH_6},\mathbf{ER}_{3}} \notag \\
&\quad \times \ket{0}_{\mathbf{BH_1},\mathbf{BH_2},\mathbf{BH_3},\mathbf{ER}_{4},\mathbf{ER}_{5},\mathbf{ER}_{6}} .
\end{align}
When tracing out the degrees of freedom, except for $\mathbf{BH_4}, \mathbf{JR}, \mathbf{ER}_{1}$, the induced density matrix for $\theta=\pi/2$ depends not only on $\gamma_1$ and $\gamma_4$, but also on $\gamma_2$ and $\gamma_3$, which gives rise to a rather different entanglement property between $\mathbf{BH}$ and $\mathbf{JR}$ than that obtained for $\theta=0$ depending only on $\gamma_1$ and $\gamma_4$ given by the above wave function.

To compute the mutual information $I(\mathbf{BH}{:}\mathbf{JR})$, we obtain the density matrix $\rho^4_{\mathbf{BH}\cup \mathbf{JR}}$ from (\ref{psi4}) by tracing out $\mathbf{ER}$ given by
\begin{equation}
\resizebox{8.5cm}{!}{$
\begin{split}
    \rho^4_{\mathbf{BH}\cup \mathbf{JR}} &= \operatorname{Tr}_{\mathbf{ER}}\!\left[ \ket{\psi^4}\bra{\psi^4} \right] \\
    &\to \left[
\begin{array}{*{16}{c}}
a^4_{1,1} & 0 & 0 & 0 & a^4_{1,10} & 0 & 0 & 0  \\
0 & a^4_{3,3} & 0 & 0 & 0 & a^4_{3,12} & 0 & 0  \\
0 & 0 & a^4_{5,5} & 0 & 0 & 0 & a^4_{5,14} & 0  \\
0 & 0 & 0 & a^4_{7,7} & 0 & 0 & 0 & a^4_{7,16}  \\
a^4_{10,1} & 0 & 0 & 0 & a^4_{10,10} & 0 & 0 & 0  \\
0 & a^4_{12,3} & 0 & 0 & 0 & a^4_{12,12} & 0 & 0 \\
0 & 0 & a^4_{14,5} & 0 & 0 & 0 & a^4_{14,14} & 0  \\
0 & 0 & 0 & a^4_{16,7} & 0 & 0 & 0 & a^4_{16,16}
\end{array}
    \right]
\end{split}
$}
\end{equation}
with the matrix elements
\begin{equation}
\resizebox{8.5cm}{!}{$
\begin{split}
    a^4_{1,1} &= a^4_{7,7} = a^4_{12,12} = a^4_{14,14} =
    \frac{1}{8} \left[ 1 + \cos\left(2\left(\gamma_1+\gamma_2+\gamma_3\right)\right)\cos{(2\gamma_4)} \right], \\
    a^4_{3,3} &= a^4_{5,5} = a^4_{10,10} = a^4_{16,16} =
    \frac{1}{8} \left[ 1 -\cos\left(2\left(\gamma_1+\gamma_2+\gamma_3\right)\right)\cos{(2\gamma_4)} \right], \\
    a^4_{1,10} &= a^4_{10,1} = a^4_{7,16} = a^4_{16,7} =
    \frac{1}{8} \left[ \cos\left(2\left(\gamma_1+\gamma_2+\gamma_3\right)\right)\sin{(2\gamma_4)} \right], \\
    a^4_{3,12} &= a^4_{5,14} = a^4_{14,5} = a^4_{12,3} = -
    \frac{1}{8} \left[ \cos\left(2\left(\gamma_1+\gamma_2+\gamma_3\right)\right)\sin{(2\gamma_4)} \right], \label{a^4_ij_pi/2}
\end{split}
$}
\end{equation}
where the matrix and its elements can also be obtained from (\ref{B1}) and (\ref{a^4_ij}) for $\theta = \pi/2$.
Some columns and rows from the original $16 \times 16$ matrix for the 4-qubit system are removed, leading to the zero eigenvalue.
Details of the matrix are shown in \ref{AppB_4}.
The eight nonzero eigenvalues are listed as
\begin{align}
\lambda_{{\mathbf{BH}\cup \mathbf{JR}}{1-4}}
&= \frac{1}{4} \cos^2\left(\gamma_1+\gamma_2+\gamma_3\right), \\
\lambda_{{\mathbf{BH}\cup \mathbf{JR}}{5-8}}
&= \frac{1}{4} \sin^2\left(\gamma_1+\gamma_2+\gamma_3\right).
\end{align}
Substituting the eigenvalues into (\ref{S}) gives the entropy $S_{\mathbf{BH}\cup \mathbf{JR}}$.
In addition, the density matrices $\rho^4_{\mathbf{BH}}$ and $\rho^4_{\mathbf{JR}}$ by tracing out $\mathbf{ER}$, $\mathbf{JR}$, and $\mathbf{ER}$, $\mathbf{BH}$, respectively, are expressed below.
The reduced density matrix $\rho^4_{\mathbf{BH}}$ is given by
\begin{equation}
\resizebox{8.5cm}{!}{$
\begin{split}
    \rho^4_{\mathbf{BH}} &=
    \operatorname{Tr}_{\mathbf{JR}}\!\left[ \rho^4_{\mathbf{BH}\cup \mathbf{JR}} \right] \\
    &= \text{diag}\left(a^4_{1,1}, a^4_{3,3}, a^4_{5,5}, a^4_{7,7}, a^4_{10,10}, a^4_{12,12}, a^4_{14,14}, a^4_{16,16} \right)
\end{split}
$}
\end{equation}
with the matrix elements denoted in (\ref{a^4_ij_pi/2}).
%
Again, part of the $16 \times 16 $ matrix for the 4-qubit system, giving zero eigenvalue, is removed.
Then the eigenvalues of the already diagonal density matrix are
\begin{align}
    \lambda_{{\mathbf{BH}}{1-4}}
    &= \frac{1}{8} \left[ 1 + \cos\left(2\left(\gamma_1+\gamma_2+\gamma_3\right)\right)\cos{(2\gamma_4)} \right], \\
    \lambda_{{\mathbf{BH}}{5-8}}
    &= \frac{1}{8} \left[ 1 - \cos\left(2\left(\gamma_1+\gamma_2+\gamma_3\right)\right)\cos{(2\gamma_4)} \right].
\end{align}
Finally, $\mathbf{JR}$ has a 1-qubit with a $2\times 2 $ density matrix given by
\begin{align}
    \rho^4_{\mathbf{JR}} &=
    { \operatorname{Tr}_{\mathbf{BH}}\!\left[ \rho^4_{\mathbf{BH}\cup \mathbf{JR}} \right] }
    = \frac{1}{2} \mathbb{I}_2 ,
\end{align}
and the eigenvalues are
\begin{align}
    \lambda_{{\mathbf{JR}}{1-2}} &= \frac{1}{2}.
\end{align}
Using (\ref{I}), one can analytically compute the mutual information $I$ as a function of $M\omega$.
In the limits of $M_4 \omega \ll 0.1$ and $M_4 \omega \gg 0.1$, $I(\mathbf{BH}:\mathbf{JR})$ has the behavior of
\begin{equation}
\resizebox{8.5cm}{!}{$
\begin{split}
    &I(\mathbf{BH}:\mathbf{JR}) \\
    &\sim\begin{cases}
    1 + \frac{8}{9 \log 2}\left(3 \sqrt{2}+2 \sqrt{3}+\sqrt{15}\right)^2 \pi^2 (M_4 \omega)^2, & M_4 \omega \ll 0.1 \\
    1 + \frac{8}{\log 2} \pi (M_4 \omega) e^{-8\pi M_4 \omega}, & M_4 \omega \gg 0.1
    \end{cases}.
\end{split}
$}
\end{equation}
which is consistent with Fig. \ref{figI}(c1).
The mutual information $I(\mathbf{BH}:\mathbf{JR})$, starting from zero, increases and then decreases to some finite value at step 4.
As seen in Fig. \ref{figI}(b1) and (c1), $I(\mathbf{BH}:\mathbf{JR})$ will decrease to zero at a later step before final step 7.
The information between $\mathbf{BH}$ and $\mathbf{JR}$ is finally lost.
The entanglement between them is also broken, resulting in the formation of the firewall, as will be seen from the negativity calculation.
This is different from their behavior at step 4 when $\theta=0$ in \cite{tok_2018} given by
\begin{align}
    I(\mathbf{BH}:\mathbf{JR}) &\sim
    \begin{cases}
    1 + \frac{16}{\log 2} \pi^2 (M_4 \omega)^2, & M_4 \omega \ll 0.1 \\
    \frac{16}{\log 2} \pi (M_4 \omega) e^{-8\pi M_4 \omega}, & M_4 \omega \gg 0.1
    \end{cases}.
\end{align}
as seen in Fig. \ref{figI}(a1).
In particular, for $M_4 \omega \gg 0.1$, which leads to insufficient entanglement in this model due to (\ref{gamma}), the mutual information between $\mathbf{BH}$ and $\mathbf{JR}$ is never established.
This marks a difference from the case of maximum scrambling at $\theta=\pi/2$, where even weak entanglement will cause $I(\mathbf{BH}:\mathbf{JR})$ to become active and finally reach zero.
For $M_4 \omega \ll 0.1$, the mutual information remains finite and will vanish at the later step with the accompanying formation of the firewall.

At this point, we study the negativity $N$ of an entanglement measure.
We take the partial transpose on $\mathbf{JR}$ denoted by $\rho^{4,T_{\mathbf{JR}}}_{\mathbf{BH},\mathbf{JR}}$, which is a $16\times 16$ matrix given by
\begin{equation}
\resizebox{9cm}{!}{$
\begin{split}
    &\rho^{4,T_{\mathbf{JR}}}_{\mathbf{BH},\mathbf{JR}} = \operatorname{Tr}^{4,T_{\mathbf{JR}}}_{\mathbf{ER}}\!\left[ \ket{\psi^4}\bra{\psi^4} \right]
    =\\
    &
    \left[
\begin{array}{*{16}{c}}
a^4_{1,1} & 0 & 0 & 0 & 0 & 0 & 0 & 0 & 0 & 0 & 0 & 0 & 0 & 0 & 0 & 0 \\
0 & 0 & 0 & 0 & 0 & 0 & 0 & 0 & a^4_{1,10} & 0 & 0 & 0 & 0 & 0 & 0 & 0 \\
0 & 0 & a^4_{3,3} & 0 & 0 & 0 & 0 & 0 & 0 & 0 & 0 & 0 & 0 & 0 & 0 & 0 \\
0 & 0 & 0 & 0 & 0 & 0 & 0 & 0 & 0 & 0 & a^4_{3,12} & 0 & 0 & 0 & 0 & 0 \\
0 & 0 & 0 & 0 & a^4_{5,5} & 0 & 0 & 0 & 0 & 0 & 0 & 0 & 0 & 0 & 0 & 0 \\
0 & 0 & 0 & 0 & 0 & 0 & 0 & 0 & 0 & 0 & 0 & 0 & a^4_{5,14} & 0 & 0 & 0 \\
0 & 0 & 0 & 0 & 0 & 0 & a^4_{7,7} & 0 & 0 & 0 & 0 & 0 & 0 & 0 & 0 & 0 \\
0 & 0 & 0 & 0 & 0 & 0 & 0 & 0 & 0 & 0 & 0 & 0 & 0 & 0 & a^4_{7,16} & 0 \\
0 & a^4_{10,1} & 0 & 0 & 0 & 0 & 0 & 0 & 0 & 0 & 0 & 0 & 0 & 0 & 0 & 0 \\
0 & 0 & 0 & 0 & 0 & 0 & 0 & 0 & 0 & a^4_{10,10} & 0 & 0 & 0 & 0 & 0 & 0 \\
0 & 0 & 0 & a^4_{12,3} & 0 & 0 & 0 & 0 & 0 & 0 & 0 & 0 & 0 & 0 & 0 & 0 \\
0 & 0 & 0 & 0 & 0 & 0 & 0 & 0 & 0 & 0 & 0 & a^4_{12,12} & 0 & 0 & 0 & 0 \\
0 & 0 & 0 & 0 & 0 & a^4_{14,5} & 0 & 0 & 0 & 0 & 0 & 0 & 0 & 0 & 0 & 0 \\
0 & 0 & 0 & 0 & 0 & 0 & 0 & 0 & 0 & 0 & 0 & 0 & 0 & a^4_{14,14} & 0 & 0 \\
0 & 0 & 0 & 0 & 0 & 0 & 0 & a^4_{16,7} & 0 & 0 & 0 & 0 & 0 & 0 & 0 & 0 \\
0 & 0 & 0 & 0 & 0 & 0 & 0 & 0 & 0 & 0 & 0 & 0 & 0 & 0 & 0 & a^4_{16,16}
\end{array}
    \right]
\end{split}
$}
\end{equation}
with the same elements as in (\ref{a^4_ij_pi/2}).
%
The 16 eigenvalues are
\begin{align}
    \lambda^T_{{\mathbf{BH},\mathbf{JR}}{1-4}}
    &= \frac{1}{8} \left[ 1 + \cos\left(2\left(\gamma_1+\gamma_2+\gamma_3\right)\right)\cos{(2\gamma_4)} \right], \\
    \lambda^T_{{\mathbf{BH},\mathbf{JR}}{5-8}}
    &= \frac{1}{8} \left[ 1 - \cos\left(2\left(\gamma_1+\gamma_2+\gamma_3\right)\right)\cos{(2\gamma_4)} \right], \\
    \lambda^T_{{\mathbf{BH},\mathbf{JR}}{9-12}}
    &= \frac{1}{8} {\left|\cos\left(2\left(\gamma_1+\gamma_2+\gamma_3\right)\right) \right|}\sin{(2\gamma_4)}, \\
    \lambda^T_{{\mathbf{BH},\mathbf{JR}}{13-16}}
    &= -\frac{1}{8} {\left|\cos\left(2\left(\gamma_1+\gamma_2+\gamma_3\right)\right) \right|}\sin{(2\gamma_4)}.
\end{align}
Using (\ref{N}), we can obtain the negativity
\begin{align}
    N(\mathbf{BH}:\mathbf{JR})=\frac{1}{2}\cos\left(2\left(\gamma_1+\gamma_2+\gamma_3\right)\right)\sin{(2\gamma_4)}
\end{align}
as a function of $M\omega$.
For the large and small values of $M_4 \omega$ with $\theta=\pi/2$, it behaves like
\begin{equation}
\resizebox{9cm}{!}{$
\begin{split}
    N(\mathbf{BH}:\mathbf{JR}) &\sim
    \begin{cases}
    \frac{2}{3} \left(3 \sqrt{2}+2 \sqrt{3}+\sqrt{15}\right)\pi (M_4 \omega), & M_4 \omega \ll 0.1 \\
    e^{-4 \pi M_4 \omega}, & M_4 \omega \gg 0.1
    \end{cases}.
\end{split}
$}
\end{equation}
The results show a firewall structure between $\mathbf{BH}$ and $\mathbf{JR}$ for various values of $M_4 \omega$, provided that $N(\mathbf{BH}:\mathbf{JR})$ is small at step 4 in Fig. \ref{figN}(c1).
As for $\theta=0$,
\begin{align}
    N(\mathbf{BH}:\mathbf{JR}) &\sim
    \begin{cases}
    2\sqrt{2} \pi \left(M_4 \omega\right), & M_4 \omega \ll 0.1 \\
    e^{-4 M_4 \omega}, & M_4 \omega \gg 0.1
    \end{cases}.
\end{align}
For $M_4 \omega \ll 0.1$, the smallness of $N(\mathbf{BH}:\mathbf{JR})$ shows the emergence of the firewall.
For $M_4 \omega \gg 0.1$, $N(\mathbf{BH}:\mathbf{JR})$ still remain zero from step 1 to step 7 due to insufficient entanglement.

\subsubsection{At step 5 for $\theta=\pi/2$}
From (\ref{psi5}), the reduced density matrix $\rho^5_{\mathbf{BH}\cup \mathbf{JR}}$ at step 5, giving the nonzero eigenvalues, now becomes
\begin{align}
    \rho^5_{\mathbf{BH}\cup \mathbf{JR}} &= \operatorname{Tr}_{\mathbf{ER}}\!\left[ \ket{\psi^5}\bra{\psi^5} \right] \notag\\
    &= \left[
\begin{array}{*{16}{c}}
a^5_{1,1} & 0 & 0 & 0 & 0 & a^5_{1,6} & 0 & 0  \\
0 & a^5_{2,2} & 0 & 0 & a^5_{2,5} & 0 & 0 & 0  \\
0 & 0 & a^5_{3,3} & 0 & 0 & 0 & 0 & a^5_{3,8}  \\
0 & 0 & 0 & a^5_{4,4} & 0 & 0 & a^5_{4,7} & 0  \\
0 & a^5_{5,2} & 0 & 0 & a^5_{5,5} & 0 & 0 & 0  \\
a^5_{6,1} & 0 & 0 & 0 & 0 & a^5_{6,6} & 0 & 0  \\
0 & 0 & 0 & a^5_{7,4} & 0 & 0 & a^5_{7,7} & 0  \\
0 & 0 & a^5_{8,3} & 0 & 0 & 0 & 0 & a^5_{8,8}
\end{array}
    \right],
\end{align}
where
\begin{align}
    a^5_{1,1} &= a^5_{4,4} = a^5_{5,5} = a^5_{8,8} \notag \\
    &=
    \frac{1}{8} \left[ 1 + \cos\left(2\left(\gamma_1+\gamma_2+\gamma_3\right)\right)\cos{(2\gamma_4)}\cos{(2\gamma_5)} \right], \notag \\
    a^5_{2,2} &= a^5_{3,3} = a^5_{6,6} = a^5_{7,7} \notag \\
    &=
    \frac{1}{8} \left[ 1 - \cos\left(2\left(\gamma_1+\gamma_2+\gamma_3\right)\right) \cos{(2\gamma_4)}\cos{(2\gamma_5)} \right], \notag \\
    a^5_{6,1} &= a^5_{1,6} = a^5_{7,4} = a^5_{4,7} \notag \\
    &=
    \frac{1}{8} \left[ \cos\left(2\left(\gamma_1+\gamma_2+\gamma_3\right)\right) \cos{(2\gamma_4)}\sin{(2\gamma_5)} \right], \notag \\
    a^5_{5,2} &= a^5_{2,5} = a^5_{8,3} = a^5_{3,8} \notag \\
    &=
    -\frac{1}{8} \left[ \cos\left(2\left(\gamma_1+\gamma_2+\gamma_3\right)\right) \cos{(2\gamma_4)}\sin{(2\gamma_5)} \right]. \label{a^5_ij_pi/2}
\end{align}
Details of the matrix are shown in \ref{AppB_5}.
The eigenvalues are
\begin{align}
    \lambda_{{\mathbf{BH}\cup \mathbf{JR}}{1-2}}
    &= \frac{1}{8} \left[ 1 + \cos\left(2\left(\gamma_1+\gamma_2+\gamma_3\right)\right)\cos{(2\gamma_4)} \right], \\
    \lambda_{{\mathbf{BH}\cup \mathbf{JR}}{3-4}}
    &= \frac{1}{8} \left[ 1 - \cos\left(2\left(\gamma_1+\gamma_2+\gamma_3\right)\right)\cos{(2\gamma_4)} \right], \\
    \lambda_{{\mathbf{BH}\cup \mathbf{JR}}{5-6}}
    &= \frac{1}{8} \left[ 1 + {\left|\cos\left(2\left(\gamma_1+\gamma_2+\gamma_3\right)\right) \right|}\cos{(2\gamma_4)} \right], \\
    \lambda_{{\mathbf{BH}\cup \mathbf{JR}}{7-8}}
    &= \frac{1}{8} \left[ 1 - {\left|\cos\left(2\left(\gamma_1+\gamma_2+\gamma_3\right)\right) \right|}\cos{(2\gamma_4)} \right].
\end{align}
The density matrix $\rho^5_{\mathbf{BH}}$ is obtained as
\begin{align}
    \rho^5_{\mathbf{BH}} &= \operatorname{Tr}_{\mathbf{JR}}\!\left[ \rho^5_{\mathbf{BH}\cup \mathbf{JR}} \right]
    = \frac{1}{4} \mathbb{I}_4 ,
\end{align}
%
and the eigenvalues are
\begin{align}
    \lambda_{{\mathbf{BH}}{1-4}} &= \frac{1}{4} .
\end{align}
Also, the density of matrix $\rho^5_{\mathbf{JR}}$ is obtained as
\begin{align}
    \rho^5_{\mathbf{JR}} &=
    { \operatorname{Tr}_{\mathbf{BH}}\!\left[ \rho^5_{\mathbf{BH}\cup \mathbf{JR}} \right] }
    = \frac{1}{2} \mathbb{I}_2 ,
\end{align}
%
and the eigenvalues are
\begin{align}
    \lambda_{{\mathbf{JR}}{1-2}} &= \frac{1}{2} .
\end{align}
The mutual information $I(\mathbf{BH}:\mathbf{JR})$ in (\ref{I}) for $\theta = \pi/2$ at step 5 behaves as follows
\begin{equation}
\resizebox{8.7cm}{!}{$
\begin{split}
    &I(\mathbf{BH}:\mathbf{JR}) \\
    &\sim
    \begin{cases}
    \frac{64}{3 \log 2} \left(4 \sqrt{5}+4 \sqrt{6} +2 \sqrt{30}+15\right) \pi^4 (M_4 \omega)^4 , & M_4 \omega \ll 0.1 \\
    1 - \frac{8}{\log 2} \pi (M_4 \omega) e^{-8\pi M_4 \omega}, & M_4 \omega \gg 0.1 \, .
    \end{cases}.
\end{split}
$}
\end{equation}
They all keep decreasing toward zero as seen in Fig. \ref{figI}(b1).
Regarding $\theta=0$, $I(\mathbf{BH}:\mathbf{JR})$ behaves as
\begin{equation}
\resizebox{8.5cm}{!}{$
\begin{split}
    &I(\mathbf{BH}:\mathbf{JR}) \\
    &\sim
    \begin{cases}
    \frac{10}{\log 2} \pi^2 (M_4 \omega)^2 , & M_4 \omega \ll 0.1 \\
    \frac{8}{\log 2} (1+\sqrt{6}) \pi (M_4 \omega) e^{-8\sqrt{\frac{2}{3}} \pi M_4 \omega}, & M_4 \omega \gg 0.1
    \end{cases}.
\end{split}
$}
\end{equation}
The mutual information $I(\mathbf{BH}:\mathbf{JR})$ for $M_4 \omega \ll 0.1$ decreases to zero at step 5, whereas it remains zero from step 1 to step 7 when $M_4 \omega \gg 0.1 $.

Similarly to step 4, we construct the partial transpose on $\mathbf{JR}$ from $\rho^{5}_{\mathbf{BH}\cup \mathbf{JR}}$.
The partially transposed density matrix $\rho^{5,T_{\mathbf{JR}}}_{\mathbf{BH},\mathbf{JR}}$ is shown as
\begin{align}
    \rho^{5,T_{\mathbf{JR}}}_{\mathbf{BH},\mathbf{JR}} &= \operatorname{Tr}^{5,T_{\mathbf{JR}}}_{\mathbf{ER}}\!\left[ \ket{\psi^5}\bra{\psi^5} \right] \notag \\
    &= \left[
\begin{array}{*{16}{c}}
a^5_{1,1} & 0 & 0 & 0 & 0 & a^5_{2,5} & 0 & 0  \\
0 & a^5_{2,2} & 0 & 0 & a^5_{1,6} & 0 & 0 & 0  \\
0 & 0 & a^5_{3,3} & 0 & 0 & 0 & 0 & a^5_{4,7}  \\
0 & 0 & 0 & a^5_{4,4} & 0 & 0 & a^5_{3,8} & 0  \\
0 & a^5_{6,1} & 0 & 0 & a^5_{5,5} & 0 & 0 & 0  \\
a^5_{5,2} & 0 & 0 & 0 & 0 & a^5_{6,6} & 0 & 0  \\
0 & 0 & 0 & a^5_{8,3} & 0 & 0 & a^5_{7,7} & 0  \\
0 & 0 & a^5_{7,4} & 0 & 0 & 0 & 0 & a^5_{8,8}
\end{array}
    \right],
\end{align}
with the same elements in (\ref{a^5_ij_pi/2}).
%
%
The eigenvalues are
\begin{align}
    \lambda^T_{{\mathbf{BH},\mathbf{JR}}{1-4}}
    &= \frac{1}{8} \left[ 1 + {\left|\cos\left(2\left(\gamma_1+\gamma_2+\gamma_3\right)\right) \right|} \cos{(2\gamma_4)} \right], \\
    \lambda^T_{{\mathbf{BH},\mathbf{JR}}{5-8}}
    &= \frac{1}{8} \left[ 1 - {\left|\cos\left(2\left(\gamma_1+\gamma_2+\gamma_3\right)\right) \right|}\cos{(2\gamma_4)} \right].
\end{align}
The negativity $N(\mathbf{BH}:\mathbf{JR})$ in (\ref{N}) for $\theta = \pi/2$ at step 5 is exactly zero for all values of $M\omega$ since all eigenvalues are positive in Fig. \ref{figN}(c1).
This can also be understood from the reduced density matrix $\rho^{5}_{\mathbf{BH5},\mathbf{JR}}$, which is already diagonal below
\begin{align}
    \rho^{5}_{\mathbf{BH5},\mathbf{JR}} &= \operatorname{Tr}^{5}_{\mathbf{ER}}\!\left[ \ket{\psi^5}\bra{\psi^5} \right]
    = \frac{1}{4} \mathbb{I}_4
\end{align}
with all positive eigenvalues, leading to $N(\mathbf{BH}:\mathbf{JR})=0$.
As for $\theta = 0$ at step 5,
\begin{align}
    N(\mathbf{BH}:\mathbf{JR}) &\sim
    \begin{cases}
    0, & M_4 \omega \ll 0.1 \\
    e^{-4 \sqrt{\frac{2}{3}} \pi M_4 \omega}. & M_4 \omega \gg 0.1
    \end{cases}.
\end{align}
All eigenvalues are positive for $M_4 \omega \ll 0.1$ giving $N(\mathbf{BH}:\mathbf{JR})=0$, but two eigenvalues are negative for $M_4 \omega \gg 0.1$, where $N(\mathbf{BH}:\mathbf{JR})$ remains small in Fig. \ref{figN}(a1).

Note that for the  initial state with $\theta=\pi/2$, a firewall emerges for all values $M \omega$ at the earlier step, compared to the case of $\theta=0$.
For a given $M_4 \omega$, the range of $\theta$, which leads to vanishing $N(\mathbf{BH}:\mathbf{JR})$, is seen at step 6 of Fig. \ref{figNtheta}.
In particular, in the limit $M_4 \omega \gg 0.1$, one of the eigenvalues of $\rho^{6,T_{\mathbf{JR}}}_{\mathbf{BH},\mathbf{JR}}$ in (\ref{rho6T}), which might change from the negative to the positive value when $\theta$ is altered, is approximated by
\begin{align}
    {\lambda_{{\mathbf{BH}},{\mathbf{JR}}}^{T}} \approx \frac{1}{8} \left( 1-\cos{2\theta} -8\cos^2{\theta} \sqrt{\sin ^2{\gamma_6} \cos ^2{\gamma_6}} \right)\, .
\end{align}
Therefore, the negativity $N(\mathbf{BH}:\mathbf{JR})$ vanishes as long as ${\lambda_{{\mathbf{BH}}, {\mathbf{JR}}}^{T}} \ge 0$.
Given a value of $M_4 \omega$, the corresponding $\theta$, which gives rise to a vanishing $N(\mathbf{BH}:\mathbf{JR})$, is restricted to the following range:
\begin{small}
\begin{align} \label{theta}
    \frac{1}{2} &\cos^{-1}{\left( \frac{3 -8|\cos{\gamma_6} \sin{\gamma_6}| -2\cos{4\gamma_6}}{ 2\cos{4\gamma_6} -1} \right)} \le
    \theta(\gamma_6(M_4 \omega)) \notag \\
    &\le
    \pi- \frac{1}{2} \cos^{-1}{\left( \frac{3 -8|\cos{\gamma_6} \sin{\gamma_6}| -2\cos{4\gamma_6}}{ 2\cos{4\gamma_6} -1} \right)} ,
\end{align}
\end{small}
where the firewall structure forms between $\mathbf{BH}$ and $\mathbf{JR}$.
In particular, when $M_4 \omega \ll 0.1$, the firewall can form with small $N(\mathbf{BH}:\mathbf{JR})$ for all ranges of $\theta \in [0,\pi]$ seen in Fig. \ref{figNtheta}.
We also would like to highlight this interesting result in this note.

At step 7, $\mathbf{BH}$ evaporates completely, meaning that $I(\mathbf{BH}:\mathbf{JR}) =0$ and $N(\mathbf{BH}:\mathbf{JR}) =0$.
All information is encoded between $\mathbf{JR}$ and $\mathbf{ER}$ with the state $\ket{\psi^7}_{\mathbf{BH},\mathbf{JR},\mathbf{ER}}$ in \ref{appendixA} given by (\ref{psi7})
with the $\theta$-dependent coefficients shown in Table \ref{Tb3}.
It may be worth mentioning that following the unitary gate dynamics, the initial black hole qubit state can be retrieved from its imprint on the final radiation state, which was originally hidden behind the black hole's horizon.

\section{Concluding remarks}
\label{sec4}
In this note, we revisit the quantum circuit model of black hole evaporation proposed in \cite{tok_2018}.
The tripartite model incorporates the systems $\mathbf{BH}$, $\mathbf{JR}$, and $\mathbf{ER}$.
To explore how quantum monogamy plays a role in determining whether a firewall can emerge between $\mathbf{BH}$ and radiation, we apply a more general initial black hole qubit state by using the scrambling unitary matrix with a parameter $\theta$ to the ground state of the qubits in matter infalling toward the black hole.
The scrambling unitary matrix reduces to no scrambling and maximum scrambling at $\theta=0$ and $\theta=\pi/2$, respectively.
In this model, the entanglement structure depends on the black hole mass $M$ and the frequency of the Hawking radiation $\omega$.
For the initial state with $\theta = \pi/2$ and for all ranges of $M\omega$, during the early stages, $I(\mathbf{BH}:\mathbf{JR})$ and $I(\mathbf{BH}:\mathbf{ER})$ increase from zero and reach their respective maximum values at the Page time.
They then decrease to zero, at which point $I(\mathbf{JR}:\mathbf{ER})$ begins to increase to a finite value.
The plots of negativity $N(\mathbf{BH}:\mathbf{JR})$, $N(\mathbf{JR}:\mathbf{ER})$, and $N(\mathbf{BH}:\mathbf{ER})$ show the firewall structure emerging between $\mathbf{BH}$ and $\mathbf{JR}$, with information being carried away by radiation.
These findings are compared with those for $\theta= 0$.
We also find that a firewall structure emerges between the black hole ($\mathbf{BH}$) and the radiation ($\mathbf{JR}$), and that the information is carried away by radiation for all values of $M\omega$, provided that $\theta$ lies within a certain analytically determined range.
This note provides a detailed analysis of tripartite entanglement arising from a one-parameter-dependent initial state when constructing a quantum circuit model for black hole evaporation.
Following the unitary gate dynamics, the $\theta$ dependence of the initial black hole qubit state can be retrieved from its imprint on the final radiation state hidden behind the horizon given by the infalling matter to the black hole. 
A firewall forms due to the disentanglement of horizon pairs according to these circuit-based toy models. However, understanding whether a firewall can exist at the event horizon for an infalling observer requires a thorough gravitational description. 

Based on the theoretical foundations established in this study, the next phase will systematically validate the proposed circuit model's viability through the following stages. First, we will use the Qiskit platform to perform noiseless state vector simulations. Next, we will integrate empirical parameters via Qiskit noise models to execute noisy classical simulations. This will allow us to characterize the evolution of quantum states under realistic decoherence and perturbations. Finally, if sufficient computational resources are available, the model will be executed on physical quantum processors to analyze discrepancies between the simulated analysis and the hardware data.

\section*{Acknowledgement}
DSL would like to thank Chen-Pin Yeh and Shoichi Kawamoto for their comments on this manuscript.
This work was supported in part by the National Science and Technology Council (NSTC) of Taiwan, Republic of China.

\appendix
\section{States at each step}
\label{appendixA}
With the notations defined here
\begin{align}
\sigma_1 &:= u_{1,-} \sin{\gamma_1} \sin{\gamma_2} + u_{2,+} \cos{\gamma_1} \cos{\gamma_2} \;,\\
\sigma_2 &:= -u_{1,-} \cos{\gamma_2} \sin{\gamma_1} + u_{2,+} \cos{\gamma_1} \sin{\gamma_2} \;,\\
\sigma_3 &:= -u_{1,-} \cos{\gamma_1} \sin{\gamma_2} + u_{2,+} \cos{\gamma_2} \sin{\gamma_1} \;,\\
\sigma_4 &:= u_{1,-} \cos{\gamma_1} \cos{\gamma_2} + u_{2,+} \sin{\gamma_1} \sin{\gamma_2} \;,\\
\sigma_5 &:= u_{1,-} \sin{\gamma_1} \sin(\gamma_2 + \gamma_3) \notag \\
&\,\quad+ \cos{\gamma_1} \Big(
u_{1,-} \sin{\gamma_2} \sin{\gamma_3} + u_{2,+} \cos{\gamma_2} \cos{\gamma_3} \Big) \;,\\
\sigma_6 &:= -u_{1,-} \cos(\gamma_1 + \gamma_3) \sin{\gamma_2} \notag \\
&\,\quad+ \cos{\gamma_2} \Big(
- u_{1,-} \cos{\gamma_3} \sin{\gamma_1} + u_{2,+} \cos{\gamma_1} \sin{\gamma_3} \Big) \;,\\
\sigma_7 &:= u_{1,-} \sin{\gamma_1} \sin{\gamma_2} \sin{\gamma_3} - u_{1,-} \cos{\gamma_2} \sin(\gamma_1 + \gamma_3) \notag \\
&\,\quad+ u_{2,+} \cos{\gamma_1} \cos{\gamma_3} \sin{\gamma_2} \;,\\
\sigma_8 &:= -u_{1,-} \sin{\gamma_1} \sin(\gamma_2 + \gamma_3) \notag \\
&\,\quad+ \cos{\gamma_1} \Big(
u_{1,-} \cos{\gamma_2} \cos{\gamma_3} + u_{2,+} \sin{\gamma_2} \sin{\gamma_3} \Big) \;,\\
\sigma_9 &:= -u_{1,-} \cos(\gamma_1 + \gamma_3) \sin{\gamma_2} \notag \\
&\,\quad+ \cos{\gamma_2} \Big(
- u_{1,-} \cos{\gamma_1} \sin{\gamma_3} + u_{2,+} \cos{\gamma_3} \sin{\gamma_1} \Big) \;,\\
\sigma_{10} &:= u_{1,-} \cos{\gamma_1} \cos(\gamma_2 + \gamma_3) \notag \\
&\,\quad+ \sin{\gamma_1} \Big(
- u_{1,-} \cos{\gamma_3} \sin{\gamma_2} + u_{2,+} \cos{\gamma_2} \sin{\gamma_3} \Big) \;,\\
\sigma_{11} &:= u_{1,-} \cos{\gamma_1} \cos(\gamma_2 + \gamma_3) \notag \\
&\,\quad+ \sin{\gamma_1} \Big(
- u_{1,-} \cos{\gamma_2} \sin{\gamma_3} + u_{2,+} \cos{\gamma_3} \sin{\gamma_2} \Big) \;,\\
\sigma_{12} &:= u_{1,-} \cos{\gamma_1} \cos{\gamma_3} \sin{\gamma_2} + u_{1,-} \cos{\gamma_2} \sin(\gamma_1 + \gamma_3) \notag \\
&\,\quad+ u_{2,+} \sin{\gamma_1} \sin{\gamma_2} \sin{\gamma_3} \;,\\
\sigma_{13} &:= u_{1,-} \cos(\gamma_1 + \gamma_3) \sin{\gamma_2} \notag \\
&\,\quad+ \cos{\gamma_2} \Big(
u_{1,-} \cos{\gamma_1} \sin{\gamma_3} - u_{2,+} \cos{\gamma_3} \sin{\gamma_1} \Big) \;,\\
\sigma_{14} &:= u_{1,-} \cos(\gamma_1 + \gamma_3) \sin{\gamma_2} \notag \\
&\,\quad+ \cos{\gamma_2} \Big(
u_{1,-} \cos{\gamma_3} \sin{\gamma_1} - u_{2,+} \cos{\gamma_1} \sin{\gamma_3} \Big) \;,\\
\sigma_{15} &:= u_{1,-} \cos{\gamma_2} \sin(\gamma_1 + \gamma_3) \notag \\
&\,\quad- \sin{\gamma_2} \Big(
u_{1,-} \sin{\gamma_1} \sin{\gamma_3} + u_{2,+} \cos{\gamma_1} \cos{\gamma_3} \Big) \;,
\end{align}
where $u_{1,\pm}$, $u_{2,\pm}$, and $u_{3,\pm}$ depend on the scrambling angle $\theta$ in (\ref{u_ipm}) and $\gamma_n$ has the $M \omega$-dependence in (\ref{gamma}), at each step of the quantum circuit in Fig. \ref{fig:circuit}, the state $\ket{\psi^n}$ can be represented in the basis
\begin{align*}
&\ket{\mathbf{BH}_{123456},\mathbf{JR},\mathbf{ER}_{123456}} .
\end{align*}
Note that some qubits, which do not change at this step, have been omitted, and they all are in the ground state.
In \cite{tok_2018}, the initial state of the qubits in the black hole is
\begin{equation}
\ket{\rm in}_{\mathbf{BH},\mathbf{JR},\mathbf{ER}} = \ket{000000}_{\mathbf{BH}} \otimes \ket{0}_{\mathbf{JR}} \otimes \ket{000000}_{\mathbf{ER}} .
\end{equation}
After applying the scrambling unitary matrix (\ref{ScramblingMatrix}), the new initial state is $\ket{\psi^0} = U_{\rm scrb} \ket{\rm in}_{\mathbf{BH},\mathbf{JR},\mathbf{ER}_1}$.
The states at step 0–7 are computed as follows:
\begin{align}
\ket{\psi^0}&_{\mathbf{BH},\mathbf{JR},\mathbf{ER}_1} \notag \\
&=
u_{2,+} \ket{00000000} \notag \\
&\quad+ u_{1,-} \Big(
\ket{00001100} + \ket{00010100} + \ket{00011000} \Big) ,
\end{align}
\begin{align}
\ket{\psi^1}&_{\mathbf{BH},\mathbf{JR},\mathbf{ER}_1} \notag \\
&=
u_{2,+} \cos{\gamma_1} \ket{00000000} \notag \\
&\quad+ u_{1,-} \sin{\gamma_1} \Big(
\ket{00000100} + \ket{00001000} + \ket{00011110} \Big) \notag \\
&\quad+ u_{1,-} \cos{\gamma_1} \Big( \ket{00001100} - \ket{00010110} - \ket{00011010} \Big) \notag \\
&\quad+ u_{2,+} \sin{\gamma_1} \ket{00010010} \, ,
\end{align}
\begin{align}
\ket{\psi^2}&_{\mathbf{BH},\mathbf{JR},\mathbf{ER}_{12}} \notag \\
&=
\sigma_{1} \ket{000000000}
+ \sigma_{2} \ket{000010100} \notag \\ &\quad
+ \sigma_{3} \ket{000100010}
+ \sigma_{4} \ket{000110110} \notag \\
&\quad + u_{1,-} \sin(\gamma_{1} + \gamma_{2}) \Big(
    \ket{000001000} - \ket{000111110} \Big) \notag \\
&\quad - u_{1,-} \cos(\gamma_{1} + \gamma_{2}) \Big(
    \ket{000011100} + \ket{000101010} \Big) ,
\end{align}
\begin{align}
\ket{\psi^3}&_{\mathbf{BH},\mathbf{JR},\mathbf{ER}_{123}} \notag \\
&=
\sigma_{5} \ket{0000000000}
+ \sigma_{6} \ket{0000011000} \notag \\ &\quad
+ \sigma_{7} \ket{0000100010}
+ \sigma_{8} \ket{0000111010} \notag \\
&\quad + \sigma_{9} \ket{0001000100}
+ \sigma_{10} \ket{0001011100} \notag \\ &\quad
+ \sigma_{11} \ket{0001100110}
+ \sigma_{12} \ket{0001111110} \, ,
\end{align}
\begin{align}
\ket{\psi^4}&_{\mathbf{BH},\mathbf{JR},\mathbf{ER}_{1234}} \notag\\
&=
{\beta^4}_{1} \ket{00000000000}
+{\beta^4}_{2} \ket{00010010000} \notag \\ &\quad
+{\beta^4}_{3} \ket{00000001000}
+{\beta^4}_{4} \ket{00010011000} \notag \\
&\quad
+{\beta^4}_{5} \ket{00000100010}
+{\beta^4}_{6} \ket{00010110010} \notag \\ &\quad
+{\beta^4}_{7} \ket{00000101010}
+{\beta^4}_{8} \ket{00010111010} \notag \\
&\quad
+{\beta^4}_{9} \ket{00001000100}
+{\beta^4}_{10} \ket{00011010100} \notag \\ &\quad
+{\beta^4}_{10} \ket{00001001100}
+{\beta^4}_{12} \ket{00011011100} \notag \\
&\quad
+{\beta^4}_{13} \ket{00001100110}
+{\beta^4}_{14} \ket{00011110110} \notag \\ &\quad
+{\beta^4}_{15} \ket{00001101110}
+{\beta^4}_{16} \ket{00011111110} \label{psi4}
\end{align}
with the coefficients ${\beta^4}_{i}$  listed in Table \ref{Tb1},
\begin{table}
\begin{tabular}{cc}
\hline
\hline
       ${\beta^4}_{1} = \sigma_5 \cos{\gamma_{4}}$
&\quad ${\beta^4}_{2} = \sigma_5 \sin{\gamma_{4}}$ \\
\hline
 ${\beta^4}_{3} = -\sigma_{13} \sin{\gamma_{4}}$
&\quad ${\beta^4}_{4} = \sigma_{13} \cos{\gamma_{4}}$ \\
\hline
       ${\beta^4}_{5} = -\sigma_{14} \cos{\gamma_{4}}$
&\quad ${\beta^4}_{6} = -\sigma_{14} \sin{\gamma_{4}}$ \\
\hline
       ${\beta^4}_{7} = \sigma_{10} \sin{\gamma_{4}}$
&\quad ${\beta^4}_{8} = -\sigma_{10} \cos{\gamma_{4}}$  \\
\hline
 ${\beta^4}_{9} = \sigma_{7} \cos{\gamma_{4}}$
&\quad ${\beta^4}_{10} = \sigma_{7} \sin{\gamma_{4}}$ \\
\hline
       ${\beta^4}_{11} = \sigma_{11} \sin{\gamma_{4}}$
&\quad ${\beta^4}_{12} = -\sigma_{11} \cos{\gamma_{4}}$  \\
\hline
 ${\beta^4}_{13} = \sigma_{8} \cos{\gamma_{4}}$
&\quad ${\beta^4}_{14} = \sigma_{8} \sin{\gamma_{4}}$ \\
\hline
       ${\beta^4}_{15} = \sigma_{12} \sin{\gamma_{4}}$
&\quad ${\beta^4}_{16} = -\sigma_{12} \cos{\gamma_{4}}$ \\
\hline
\hline
\end{tabular}
\caption{\label{Tb1}
    The coefficients ${\beta^4}_{i}$ of the state at step 4, where $\sigma_j$ contributes to the black mass $M_{\rm in} \omega$ and the scrambling angle $\theta$; $\gamma_n$ contributes to the black mass $M_{\rm in} \omega$ }
\end{table}
%
\begin{align}
\ket{\psi^5}&_{\mathbf{BH},\mathbf{JR},\mathbf{ER}_{12345}} \notag\\
&=
{\beta^5}_{1} \ket{000000000000}
+{\beta^5}_{2} \ket{000000100010} \notag \\ &\quad
+{\beta^5}_{3}\ket{000010000010}
+{\beta^5}_{4} \ket{000010100000} \notag \\
&\quad
+{\beta^5}_{5} \ket{000000001000}
+{\beta^5}_{6} \ket{000000101010} \notag \\
&\quad
+{\beta^5}_{7} \ket{000000010000}
+{\beta^5}_{8} \ket{000000110010} \notag \\ &\quad
+{\beta^5}_{9} \ket{000010010010}
+{\beta^5}_{10} \ket{000010110000} \notag \\
&\quad
+{\beta^5}_{11} \ket{000000011000}
+{\beta^5}_{12} \ket{000000111010} \notag \\ &\quad
+{\beta^5}_{13} \ket{000010011010}
+{\beta^5}_{14} \ket{000010111000} \notag \\
&\quad
+{\beta^5}_{15} \ket{000001000100}
+{\beta^5}_{16} \ket{000001100110} \notag \\ &\quad
+{\beta^5}_{17} \ket{000011000110}
+{\beta^5}_{18} \ket{000011100100} \notag \\
&\quad
+{\beta^5}_{19} \ket{000001001100}
+{\beta^5}_{20} \ket{000001101110} \notag \\ &\quad
+{\beta^5}_{21} \ket{000011001110}
+{\beta^5}_{22} \ket{000011101100} \notag \\
&\quad
+{\beta^5}_{23} \ket{000001010100}
+{\beta^5}_{24} \ket{000001110110} \notag \\ &\quad
+{\beta^5}_{25} \ket{000011010110}
+{\beta^5}_{26} \ket{000011110100} \notag \\
&\quad
+{\beta^5}_{27} \ket{000001011100}
+{\beta^5}_{28} \ket{000001111110} \notag \\ &\quad
+{\beta^5}_{29} \ket{000011011110}
+{\beta^5}_{30} \ket{000011111100} \notag \\
&\quad
+{\beta^5}_{31} \ket{000010001010}
+{\beta^5}_{32} \ket{000010101000} \label{psi5}
\end{align}
with the coefficients ${\beta^5}_{i}$ listed in Table \ref{Tb2},
\begin{table*}
\begin{center}
\begin{tabular}{cccc}
\hline
\hline
       ${\beta^5}_{1} = \sigma_5 \cos{\gamma_{4}}\cos{\gamma_{5}}$
&\quad ${\beta^5}_{2} = \sigma_5 \cos{\gamma_{5}}\sin{\gamma_{4}}$
&\quad ${\beta^5}_{3} = \sigma_5 \sin{\gamma_{4}}\sin{\gamma_{5}}$
&\quad ${\beta^5}_{4} = \sigma_5 \cos{\gamma_{4}}\sin{\gamma_{5}}$ \\
\hline
       ${\beta^5}_{5} = \sigma_7 \cos{\gamma_{4}}\sin{\gamma_{5}}$
&\quad ${\beta^5}_{6} = \sigma_7 \sin{\gamma_{4}}\sin{\gamma_{5}}$ \\
\hline
       ${\beta^5}_{7} = -\sigma_{13} \cos{\gamma_{5}}\sin{\gamma_{4}}$
&\quad ${\beta^5}_{8} = \sigma_{13} \cos{\gamma_{4}}\cos{\gamma_{5}}$
&\quad ${\beta^5}_{9} = \sigma_{13} \cos{\gamma_{4}}\sin{\gamma_{5}}$
&\quad ${\beta^5}_{10} = -\sigma_{13} \sin{\gamma_{4}}\sin{\gamma_{5}}$ \\
\hline
       ${\beta^5}_{11} = \sigma_{11} \sin{\gamma_{4}}\sin{\gamma_{5}}$
&\quad ${\beta^5}_{12} = -\sigma_{11} \cos{\gamma_{4}}\sin{\gamma_{5}}$
&\quad ${\beta^5}_{13} = \sigma_{11} \cos{\gamma_{4}}\cos{\gamma_{5}}$
&\quad ${\beta^5}_{14} = -\sigma_{11} \cos{\gamma_{5}}\sin{\gamma_{4}}$ \\
\hline
       ${\beta^5}_{15} = -\sigma_{14} \cos{\gamma_{4}}\cos{\gamma_{5}}$
&\quad ${\beta^5}_{16} = -\sigma_{14} \cos{\gamma_{5}}\sin{\gamma_{4}}$
&\quad ${\beta^5}_{17} = -\sigma_{14} \sin{\gamma_{4}}\sin{\gamma_{5}}$
&\quad ${\beta^5}_{18} = -\sigma_{14} \cos{\gamma_{4}}\sin{\gamma_{5}}$ \\
\hline
       ${\beta^5}_{19} = \sigma_{8} \cos{\gamma_{4}}\sin{\gamma_{5}}$
&\quad ${\beta^5}_{20} = \sigma_{8} \sin{\gamma_{4}}\sin{\gamma_{5}}$
&\quad ${\beta^5}_{21} = -\sigma_{8} \cos{\gamma_{5}}\sin{\gamma_{4}}$
&\quad ${\beta^5}_{22} = -\sigma_{8} \cos{\gamma_{4}}\cos{\gamma_{5}}$ \\
\hline
       ${\beta^5}_{23} = \sigma_{10} \cos{\gamma_{5}}\sin{\gamma_{4}}$
&\quad ${\beta^5}_{24} = -\sigma_{10} \cos{\gamma_{4}}\cos{\gamma_{5}}$
&\quad ${\beta^5}_{25} = -\sigma_{10} \cos{\gamma_{4}}\sin{\gamma_{5}}$
&\quad ${\beta^5}_{26} = \sigma_{10} \sin{\gamma_{4}}\sin{\gamma_{5}}$ \\
\hline
       ${\beta^5}_{27} = \sigma_{12} \sin{\gamma_{4}}\sin{\gamma_{5}}$
&\quad ${\beta^5}_{28} = -\sigma_{12} \cos{\gamma_{4}}\sin{\gamma_{5}}$
&\quad ${\beta^5}_{29} = \sigma_{12} \cos{\gamma_{4}}\cos{\gamma_{5}}$
&\quad ${\beta^5}_{30} = -\sigma_{12} \cos{\gamma_{5}}\sin{\gamma_{4}}$ \\
\hline
       ${\beta^5}_{31} = \sigma_{15} \cos{\gamma_{5}}\sin{\gamma_{4}}$
&\quad ${\beta^5}_{32} = \sigma_{15} \cos{\gamma_{4}}\cos{\gamma_{5}}$ \\
\hline
\hline
\end{tabular}
\end{center}
    \caption{\label{Tb2}
    The coefficients ${\beta^5}_{i}$ of the state at step 5}
\end{table*}
%
\begin{align}
\ket{\psi^6}&_{\mathbf{BH},\mathbf{JR},\mathbf{ER}} \notag \\
&=
      {\beta^6}_{1} \ket{0000000000000}
    + {\beta^6}_{2} \ket{0000000000110} \notag \\ &\quad
    + {\beta^6}_{3} \ket{0000001000010}
    + {\beta^6}_{4} \ket{0000001000100} \notag \\
&\quad
    + {\beta^6}_{5} \ket{0000010000010}
    + {\beta^6}_{6} \ket{0000010000100} \notag \\ &\quad
    + {\beta^6}_{7} \ket{0000011000000}
    + {\beta^6}_{8} \ket{0000011000110} \notag \\
&\quad
    +{\beta^6}_{9} \ket{0000000001000}
    +{\beta^6}_{10} \ket{0000000001110} \notag \\ &\quad
    +{\beta^6}_{11} \ket{0000001001010}
    +{\beta^6}_{12} \ket{0000001001100} \notag \\
&\quad
    +{\beta^6}_{13} \ket{0000010001010}
    +{\beta^6}_{14} \ket{0000010001100} \notag \\ &\quad
    +{\beta^6}_{15} \ket{0000011001000}
    +{\beta^6}_{16} \ket{0000011001110} \notag \\
&\quad
    +{\beta^6}_{17} \ket{0000000010000}
    +{\beta^6}_{18} \ket{0000000010110} \notag \\ &\quad
    +{\beta^6}_{19} \ket{0000011010000}
    +{\beta^6}_{20} \ket{0000011010110} \notag \\
&\quad
    +{\beta^6}_{21} \ket{0000000011000}
    +{\beta^6}_{22} \ket{0000000011110} \notag \\ &\quad
    +{\beta^6}_{23} \ket{0000001011010}
    +{\beta^6}_{24} \ket{0000001011100} \notag \\
&\quad
    +{\beta^6}_{25} \ket{0000010011010}
    +{\beta^6}_{26} \ket{0000010011100} \notag \\ &\quad
    +{\beta^6}_{27} \ket{0000011011000}
    +{\beta^6}_{28} \ket{0000011011110} \notag \\
&\quad
    +{\beta^6}_{29} \ket{0000000100000}
    +{\beta^6}_{30} \ket{0000000100110} \notag \\ &\quad
    +{\beta^6}_{31} \ket{0000001100010}
    +{\beta^6}_{32} \ket{0000001100100} \notag \\
&\quad
    +{\beta^6}_{33} \ket{0000010100010}
    +{\beta^6}_{34} \ket{0000010100100} \notag \\ &\quad
    +{\beta^6}_{35} \ket{0000011100000}
    +{\beta^6}_{36} \ket{0000011100110} \notag \\
&\quad
    +{\beta^6}_{37} \ket{0000000101000}
    +{\beta^6}_{38} \ket{0000000101110} \notag \\ &\quad
    +{\beta^6}_{39} \ket{0000001101010}
    +{\beta^6}_{40} \ket{0000001101100} \notag \\
&\quad
    +{\beta^6}_{41} \ket{0000010101010}
    +{\beta^6}_{42} \ket{0000010101100} \notag \\ &\quad
    +{\beta^6}_{43} \ket{0000011101000}
    +{\beta^6}_{44} \ket{0000011101110} \notag \\
&\quad
    +{\beta^6}_{45} \ket{0000000110000}
    +{\beta^6}_{46} \ket{0000000110110} \notag \\ &\quad
    +{\beta^6}_{47} \ket{0000001110010}
    +{\beta^6}_{48} \ket{0000001110100} \notag \\
&\quad
    +{\beta^6}_{49} \ket{0000010110010}
    +{\beta^6}_{50} \ket{0000010110100} \notag \\ &\quad
    +{\beta^6}_{51} \ket{0000011110000}
    +{\beta^6}_{52} \ket{0000011110110} \notag \\
&\quad
    +{\beta^6}_{53} \ket{0000000111000}
    +{\beta^6}_{54} \ket{0000000111110} \notag \\ &\quad
    +{\beta^6}_{55} \ket{0000001111010}
    +{\beta^6}_{56} \ket{0000001111100} \notag \\
&\quad
    +{\beta^6}_{57} \ket{0000010111010}
    +{\beta^6}_{58} \ket{0000010111100} \notag \\ &\quad
    +{\beta^6}_{59} \ket{0000011111000}
    +{\beta^6}_{60} \ket{0000011111110} \notag \\
&\quad
    +{\beta^6}_{61} \ket{0000001010010}
    +{\beta^6}_{62} \ket{0000001010100} \notag \\ &\quad
    +{\beta^6}_{63} \ket{0000010010010}
    +{\beta^6}_{64} \ket{0000010010100} \label{psi6}
\end{align}
with the coefficients ${\beta^6}_{i}$ listed in Table \ref{Tb3},
and
\begin{footnotesize}
\begin{align}
\ket{\psi^7}&_{\mathbf{BH},\mathbf{JR},\mathbf{ER}} \notag \\
&=
{\beta^7}_{1} \ket{0000000000000}
+{\beta^7}_{2} \ket{0000000000011} \notag \\ &\quad
+{\beta^7}_{3} \ket{0000000000101}
+{\beta^7}_{4} \ket{0000000000110} \notag \\
&\quad
+{\beta^7}_{5} \ket{0000001000001}
+{\beta^7}_{6} \ket{0000001000010} \notag \\ &\quad
+{\beta^7}_{7} \ket{0000001000100}
+{\beta^7}_{8} \ket{0000001000111} \notag \\
&\quad
+{\beta^7}_{9} \ket{0000000001000}
+{\beta^7}_{10} \ket{0000000001011} \notag \\ &\quad
+{\beta^7}_{11} \ket{0000000001101}
+{\beta^7}_{12} \ket{0000000001110} \notag \\
&\quad
+{\beta^7}_{13} \ket{0000001001001}
+{\beta^7}_{14} \ket{0000001001010} \notag \\ &\quad
+{\beta^7}_{15} \ket{0000001001100}
+{\beta^7}_{16} \ket{0000001001111} \notag \\
&\quad
+{\beta^7}_{17} \ket{0000000010000}
+{\beta^7}_{18} \ket{0000000010110} \notag \\ &\quad
+{\beta^7}_{19} \ket{0000001010001}
+{\beta^7}_{20} \ket{0000001010111} \notag \\
&\quad
+{\beta^7}_{21} \ket{0000000010011}
+{\beta^7}_{22} \ket{0000000010101} \notag \\ &\quad
+{\beta^7}_{23} \ket{0000001010010}
+{\beta^7}_{24} \ket{0000001010100} \notag \\
&\quad
+{\beta^7}_{25} \ket{0000000011000}
+{\beta^7}_{26} \ket{0000000011011} \notag \\ &\quad
+{\beta^7}_{27} \ket{0000000011101}
+{\beta^7}_{28} \ket{0000000011110} \notag \\
&\quad
+{\beta^7}_{29} \ket{0000001011001}
+{\beta^7}_{30} \ket{0000001011010} \notag \\ &\quad
+{\beta^7}_{31} \ket{0000001011100}
+{\beta^7}_{32} \ket{0000001011111} \notag \\
&\quad
+{\beta^7}_{33} \ket{0000000100000}
+{\beta^7}_{34} \ket{0000000100011} \notag \\ &\quad
+{\beta^7}_{35} \ket{0000000100101}
+{\beta^7}_{36} \ket{0000000100110} \notag \\
&\quad
+{\beta^7}_{37} \ket{0000001100001}
+{\beta^7}_{38} \ket{0000001100010} \notag \\ &\quad
+{\beta^7}_{39} \ket{0000001100100}
+{\beta^7}_{40} \ket{0000001100111} \notag \\
&\quad
+{\beta^7}_{41} \ket{0000000101000}
+{\beta^7}_{42} \ket{0000000101011} \notag \\ &\quad
+{\beta^7}_{43} \ket{0000000101101}
+{\beta^7}_{44} \ket{0000000101110} \notag \\
&\quad
+{\beta^7}_{45} \ket{0000001101001}
+{\beta^7}_{46} \ket{0000001101010} \notag \\ &\quad
+{\beta^7}_{47} \ket{0000001101100}
+{\beta^7}_{48} \ket{0000001101111} \notag \\
&\quad
+{\beta^7}_{49} \ket{0000000110000}
+{\beta^7}_{50} \ket{0000000110011} \notag \\ &\quad
+{\beta^7}_{51} \ket{0000000110101}
+{\beta^7}_{52} \ket{0000000110110} \notag \\
&\quad
+{\beta^7}_{53} \ket{0000001110001}
+{\beta^7}_{54} \ket{0000001110010} \notag \\ &\quad
+{\beta^7}_{55} \ket{0000001110100}
+{\beta^7}_{56} \ket{0000001110111} \notag \\
&\quad
+{\beta^7}_{57} \ket{0000000111000}
+{\beta^7}_{58} \ket{0000000111011} \notag \\ &\quad
+{\beta^7}_{59} \ket{0000000111101}
+{\beta^7}_{60} \ket{0000000111110} \notag \\
&\quad
+{\beta^7}_{61} \ket{0000001111001}
+{\beta^7}_{62} \ket{0000001111010} \notag \\ &\quad
+{\beta^7}_{63} \ket{0000001111100}
+{\beta^7}_{64} \ket{0000001111111} \label{psi7}
\end{align}
\end{footnotesize}
with the coefficients ${\beta^7}_{i}$, which are the rearrangement of ${\beta^6}_{i}$ shown in Table \ref{Tb3}.

\begin{table*}
\begin{center}
\begin{tabular}{ccc}
\hline
\hline
       ${\beta^6}_{1} = \sigma_5 \cos{\gamma_4}\cos{\gamma_5}\cos{\gamma_6} = {\beta^7}_{1}$
&\quad ${\beta^6}_{2} = \sigma_5 \cos{\gamma_5}\cos{\gamma_6}\sin{\gamma_4} = {\beta^7}_{4}$
&\quad ${\beta^6}_{3} = \sigma_5 \cos{\gamma_4}\cos{\gamma_6}\sin{\gamma_5} = {\beta^7}_{2}$ \\
       ${\beta^6}_{4} = \sigma_5 \cos{\gamma_6}\sin{\gamma_4}\sin{\gamma_5} = {\beta^7}_{3}$
&\quad ${\beta^6}_{5} = \sigma_5 \cos{\gamma_4}\sin{\gamma_5}\sin{\gamma_6} = {\beta^7}_{6}$
&\quad ${\beta^6}_{6} = \sigma_5 \sin{\gamma_4}\sin{\gamma_5}\sin{\gamma_6} = {\beta^7}_{7}$ \\
       ${\beta^6}_{7} = \sigma_5 \cos{\gamma_4}\cos{\gamma_5}\sin{\gamma_6} = {\beta^7}_{5}$
&\quad ${\beta^6}_{8} = \sigma_5 \cos{\gamma_5}\sin{\gamma_4}\sin{\gamma_6} = {\beta^7}_{8}$ \\
\hline
       ${\beta^6}_{9} = -\sigma_{14} \cos{\gamma_4}\cos{\gamma_5}\sin{\gamma_6} = {\beta^7}_{9}$
&\quad ${\beta^6}_{10} = -\sigma_{14} \cos{\gamma_5}\sin{\gamma_4}\sin{\gamma_6} = {\beta^7}_{12}$
&\quad ${\beta^6}_{11} = -\sigma_{14} \cos{\gamma_4}\sin{\gamma_5}\sin{\gamma_6} = {\beta^7}_{10}$ \\
       ${\beta^6}_{12} = -\sigma_{14} \sin{\gamma_4}\sin{\gamma_5}\sin{\gamma_6} = {\beta^7}_{11}$
&\quad ${\beta^6}_{13} = \sigma_{14} \cos{\gamma_4}\cos{\gamma_6}\sin{\gamma_5} = {\beta^7}_{14}$
&\quad ${\beta^6}_{14} = \sigma_{14} \cos{\gamma_6}\sin{\gamma_4}\sin{\gamma_5} = {\beta^7}_{15}$ \\
       ${\beta^6}_{15} = \sigma_{14} \cos{\gamma_4}\cos{\gamma_5}\cos{\gamma_6} = {\beta^7}_{13}$
&\quad ${\beta^6}_{16} = \sigma_{14} \cos{\gamma_5}\cos{\gamma_6}\sin{\gamma_4} = {\beta^7}_{16}$ \\
\hline
       ${\beta^6}_{17} = \sigma_{7} \cos{\gamma_4}\cos{\gamma_6}\sin{\gamma_5} = {\beta^7}_{17}$
&\quad ${\beta^6}_{18} = \sigma_{7} \cos{\gamma_6}\sin{\gamma_4}\sin{\gamma_5} = {\beta^7}_{18}$
&\quad ${\beta^6}_{19} = \sigma_{7} \cos{\gamma_4}\sin{\gamma_5}\sin{\gamma_6} = {\beta^7}_{19}$ \\
       ${\beta^6}_{20} = \sigma_{7} \sin{\gamma_4}\sin{\gamma_5}\sin{\gamma_6} = {\beta^7}_{20}$ \\
\hline
       ${\beta^6}_{21} = \sigma_{8} \cos{\gamma_4}\sin{\gamma_5}\sin{\gamma_6} = {\beta^7}_{25}$
&\quad ${\beta^6}_{22} = \sigma_{8} \sin{\gamma_4}\sin{\gamma_5}\sin{\gamma_6} = {\beta^7}_{28}$
&\quad ${\beta^6}_{23} = -\sigma_{8} \cos{\gamma_4}\cos{\gamma_5}\sin{\gamma_6} = {\beta^7}_{26}$ \\
       ${\beta^6}_{24} = -\sigma_{8} \cos{\gamma_5}\sin{\gamma_4}\sin{\gamma_6} = {\beta^7}_{27}$
&\quad ${\beta^6}_{25} = \sigma_{8} \cos{\gamma_4}\cos{\gamma_5}\cos{\gamma_6} = {\beta^7}_{30}$
&\quad ${\beta^6}_{26} = \sigma_{8} \cos{\gamma_5}\cos{\gamma_6}\sin{\gamma_4} = {\beta^7}_{31}$ \\
       ${\beta^6}_{27} = -\sigma_{8} \cos{\gamma_4}\cos{\gamma_6}\sin{\gamma_5} = {\beta^7}_{29}$
&\quad ${\beta^6}_{28} = -\sigma_{8} \cos{\gamma_6}\sin{\gamma_4}\sin{\gamma_5} = {\beta^7}_{32}$ \\
\hline
       ${\beta^6}_{29} = -\sigma_{13} \cos{\gamma_5}\cos{\gamma_6}\sin{\gamma_4} = {\beta^7}_{33}$
&\quad ${\beta^6}_{30} = \sigma_{13} \cos{\gamma_4}\cos{\gamma_5}\cos{\gamma_6} = {\beta^7}_{36}$
&\quad ${\beta^6}_{31} = -\sigma_{13} \cos{\gamma_6}\sin{\gamma_4}\sin{\gamma_5} = {\beta^7}_{34}$ \\
       ${\beta^6}_{32} = \sigma_{13} \cos{\gamma_4}\cos{\gamma_6}\sin{\gamma_5} = {\beta^7}_{35}$
&\quad ${\beta^6}_{33} = -\sigma_{13} \sin{\gamma_4}\sin{\gamma_5}\sin{\gamma_6} = {\beta^7}_{38}$
&\quad ${\beta^6}_{34} = \sigma_{13} \cos{\gamma_4}\sin{\gamma_5}\sin{\gamma_6} = {\beta^7}_{39}$ \\
       ${\beta^6}_{35} = -\sigma_{13} \cos{\gamma_5}\sin{\gamma_4}\sin{\gamma_6} = {\beta^7}_{37}$
&\quad ${\beta^6}_{36} = \sigma_{13} \cos{\gamma_4}\cos{\gamma_5}\sin{\gamma_6} = {\beta^7}_{40}$ \\
\hline
       ${\beta^6}_{37} = \sigma_{10} \cos{\gamma_5}\sin{\gamma_4}\sin{\gamma_6} = {\beta^7}_{41}$
&\quad ${\beta^6}_{38} = -\sigma_{10} \cos{\gamma_4}\cos{\gamma_5}\sin{\gamma_6} = {\beta^7}_{44}$
&\quad ${\beta^6}_{39} = \sigma_{10} \sin{\gamma_4}\sin{\gamma_5}\sin{\gamma_6} = {\beta^7}_{42}$ \\
       ${\beta^6}_{40} = -\sigma_{10} \cos{\gamma_4}\sin{\gamma_5}\sin{\gamma_6} = {\beta^7}_{43}$
&\quad ${\beta^6}_{41} = -\sigma_{10} \cos{\gamma_6}\sin{\gamma_4}\sin{\gamma_5}  = {\beta^7}_{46}$
&\quad ${\beta^6}_{42} = \sigma_{10} \cos{\gamma_4}\cos{\gamma_6}\sin{\gamma_5} = {\beta^7}_{47}$ \\
       ${\beta^6}_{43} = -\sigma_{10} \cos{\gamma_5}\cos{\gamma_6}\sin{\gamma_4} = {\beta^7}_{45}$
&\quad ${\beta^6}_{44} = \sigma_{10} \cos{\gamma_4}\cos{\gamma_5}\cos{\gamma_6} = {\beta^7}_{48}$ \\
\hline
       ${\beta^6}_{45} = \sigma_{11} \cos{\gamma_6}\sin{\gamma_4}\sin{\gamma_5} = {\beta^7}_{49}$
&\quad ${\beta^6}_{46} = -\sigma_{11} \cos{\gamma_4}\cos{\gamma_6}\sin{\gamma_5} = {\beta^7}_{52}$
&\quad ${\beta^6}_{47} = -\sigma_{11} \cos{\gamma_5}\cos{\gamma_6}\sin{\gamma_4} = {\beta^7}_{50}$ \\
       ${\beta^6}_{48} = \sigma_{11} \cos{\gamma_4}\cos{\gamma_5}\cos{\gamma_6} = {\beta^7}_{51}$
&\quad ${\beta^6}_{49} = -\sigma_{11} \cos{\gamma_5}\sin{\gamma_4}\sin{\gamma_6} = {\beta^7}_{54}$
&\quad ${\beta^6}_{50} = \sigma_{11} \cos{\gamma_4}\cos{\gamma_5}\sin{\gamma_6} = {\beta^7}_{55}$ \\
       ${\beta^6}_{51} = \sigma_{11} \sin{\gamma_4}\sin{\gamma_5}\sin{\gamma_6} = {\beta^7}_{53}$
&\quad ${\beta^6}_{52} = -\sigma_{11} \cos{\gamma_4}\sin{\gamma_5}\sin{\gamma_6} = {\beta^7}_{56}$ \\
\hline
       ${\beta^6}_{53} = \sigma_{12} \sin{\gamma_4}\sin{\gamma_5}\sin{\gamma_6} = {\beta^7}_{57}$
&\quad ${\beta^6}_{54} = -\sigma_{12} \cos{\gamma_4}\sin{\gamma_5}\sin{\gamma_6} = {\beta^7}_{60}$
&\quad ${\beta^6}_{55} = -\sigma_{12} \cos{\gamma_5}\sin{\gamma_4}\sin{\gamma_6} = {\beta^7}_{58}$ \\
       ${\beta^6}_{56} = \sigma_{12} \cos{\gamma_4}\cos{\gamma_5}\sin{\gamma_6} = {\beta^7}_{59}$
&\quad ${\beta^6}_{57} = \sigma_{12} \cos{\gamma_5}\cos{\gamma_6}\sin{\gamma_4} = {\beta^7}_{62}$
&\quad ${\beta^6}_{58} = -\sigma_{12} \cos{\gamma_4}\cos{\gamma_5}\cos{\gamma_6} = {\beta^7}_{63}$ \\
       ${\beta^6}_{59} = -\sigma_{12} \cos{\gamma_6}\sin{\gamma_4}\sin{\gamma_5} = {\beta^7}_{61}$
&\quad ${\beta^6}_{60} = \sigma_{12} \cos{\gamma_4}\cos{\gamma_6}\sin{\gamma_5} = {\beta^7}_{64}$ \\
\hline
       ${\beta^6}_{61} = \sigma_{15} \cos\gamma_4 \cos\gamma_5 \cos\gamma_6 = {\beta^7}_{21}$
&\quad ${\beta^6}_{62} = \sigma_{15} \cos\gamma_5 \cos\gamma_6 \sin\gamma_4 = {\beta^7}_{22}$
&\quad ${\beta^6}_{63} = \sigma_{15} \cos\gamma_4 \cos\gamma_5 \sin\gamma_6 = {\beta^7}_{23}$ \\
       ${\beta^6}_{64} = \sigma_{15} \cos\gamma_5 \sin\gamma_4 \sin\gamma_6 = {\beta^7}_{24}$ \\
\hline
\hline
\end{tabular}
\end{center}
    \caption{\label{Tb3}
    The coefficients ${\beta^6}_{i}$ and ${\beta^7}_{i}$ of the states at step 6 and step 7 }
\end{table*}

\section{Density matrix at each step for an arbitrary scrambling angle}
\label{appendixB}
In this section, we show the density matrices $\rho^n_{\mathbf{BH}\cup \mathbf{JR}}$, $\rho^n_{\mathbf{BH}}$, and $\rho^n_{\mathbf{JR}}$ and the partially transposed matrix $\rho^{n,T_{\mathbf{JR}}}_{\mathbf{BH},\mathbf{JR}}$, respectively, at step $n$.

\subsection{At step 4}
\label{AppB_4}
From the state in (\ref{psi4}), we can compute the density matrix of $\mathbf{BH}$ and $\mathbf{JR}$, $\rho^4_{\mathbf{BH}\cup \mathbf{JR}}$, and show in the following form
\begin{equation}
\resizebox{8.5cm}{!}{$
\begin{split}
    &\rho^4_{\mathbf{BH}\cup \mathbf{JR}} = \operatorname{Tr}_{\mathbf{ER}}\!\left[ \ket{\psi^4}\bra{\psi^4} \right] \\
    &= \left[
\begin{array}{*{16}{c}}
a^4_{1,1} & 0 & 0 & 0 & 0 & 0 & 0 & 0 & 0 & a^4_{1,10} & 0 & 0 & 0 & 0 & 0 & 0 \\
0 & 0 & 0 & 0 & 0 & 0 & 0 & 0 & 0 & 0 & 0 & 0 & 0 & 0 & 0 & 0 \\
0 & 0 & a^4_{3,3} & 0 & 0 & 0 & 0 & 0 & 0 & 0 & 0 & a^4_{3,12} & 0 & 0 & 0 & 0 \\
0 & 0 & 0 & 0 & 0 & 0 & 0 & 0 & 0 & 0 & 0 & 0 & 0 & 0 & 0 & 0 \\
0 & 0 & 0 & 0 & a^4_{5,5} & 0 & 0 & 0 & 0 & 0 & 0 & 0 & 0 & a^4_{5,14} & 0 & 0 \\
0 & 0 & 0 & 0 & 0 & 0 & 0 & 0 & 0 & 0 & 0 & 0 & 0 & 0 & 0 & 0 \\
0 & 0 & 0 & 0 & 0 & 0 & a^4_{7,7} & 0 & 0 & 0 & 0 & 0 & 0 & 0 & 0 & a^4_{7,16} \\
0 & 0 & 0 & 0 & 0 & 0 & 0 & 0 & 0 & 0 & 0 & 0 & 0 & 0 & 0 & 0 \\
0 & 0 & 0 & 0 & 0 & 0 & 0 & 0 & 0 & 0 & 0 & 0 & 0 & 0 & 0 & 0 \\
a^4_{10,1} & 0 & 0 & 0 & 0 & 0 & 0 & 0 & 0 & a^4_{10,10} & 0 & 0 & 0 & 0 & 0 & 0 \\
0 & 0 & 0 & 0 & 0 & 0 & 0 & 0 & 0 & 0 & 0 & 0 & 0 & 0 & 0 & 0 \\
0 & 0 & a^4_{12,3} & 0 & 0 & 0 & 0 & 0 & 0 & 0 & 0 & a^4_{12,12} & 0 & 0 & 0 & 0 \\
0 & 0 & 0 & 0 & 0 & 0 & 0 & 0 & 0 & 0 & 0 & 0 & 0 & 0 & 0 & 0 \\
0 & 0 & 0 & 0 & a^4_{14,5} & 0 & 0 & 0 & 0 & 0 & 0 & 0 & 0 & a^4_{14,14} & 0 & 0 \\
0 & 0 & 0 & 0 & 0 & 0 & 0 & 0 & 0 & 0 & 0 & 0 & 0 & 0 & 0 & 0 \\
0 & 0 & 0 & 0 & 0 & 0 & a^4_{16,7} & 0 & 0 & 0 & 0 & 0 & 0 & 0 & 0 & a^4_{16,16}
\end{array}
    \right] \\
    &\to \left[
\begin{array}{*{16}{c}}
a^4_{1,1} & 0 & 0 & 0 & a^4_{1,10} & 0 & 0 & 0  \\
0 & a^4_{3,3} & 0 & 0 & 0 & a^4_{3,12} & 0 & 0  \\
0 & 0 & a^4_{5,5} & 0 & 0 & 0 & a^4_{5,14} & 0  \\
0 & 0 & 0 & a^4_{7,7} & 0 & 0 & 0 & a^4_{7,16}  \\
a^4_{10,1} & 0 & 0 & 0 & a^4_{10,10} & 0 & 0 & 0  \\
0 & a^4_{12,3} & 0 & 0 & 0 & a^4_{12,12} & 0 & 0 \\
0 & 0 & a^4_{14,5} & 0 & 0 & 0 & a^4_{14,14} & 0  \\
0 & 0 & 0 & a^4_{16,7} & 0 & 0 & 0 & a^4_{16,16}
\end{array}
    \right],
\label{B1}
\end{split}
$}
\end{equation}
where some of the columns and rows giving zero eigenvalues are deleted and the matrix elements are expressed in terms of ${\beta^4}_i$ in Table \ref{Tb1},
\begin{align}
    a^4_{1,1}
    &= \left|{\beta^4}_1 \right|^2 +\left|{\beta^4}_2 \right|^2, \notag \\
    a^4_{3,3}
    &= \left|{\beta^4}_3 \right|^2 +\left|{\beta^4}_4 \right|^2 ,\notag \\
    a^4_{5,5} &= \left|{\beta^4}_5 \right|^2 +\left|{\beta^4}_6 \right|^2, \notag \\
    a^4_{7,7} &= \left|{\beta^4}_7 \right|^2 +\left|{\beta^4}_8 \right|^2 ,\notag \\
    a^4_{10,10}
    &= \left|{\beta^4}_9 \right|^2 +\left|{\beta^4}_{10} \right|^2, \notag \\
    a^4_{12,12}
    &= \left|{\beta^4}_{11} \right|^2 +\left|{\beta^4}_{12} \right|^2 ,\notag \\
    a^4_{14,14}
    &= \left|{\beta^4}_{13} \right|^2 +\left|{\beta^4}_{14} \right|^2, \notag \\
    a^4_{16,16}
    &= \left|{\beta^4}_{15} \right|^2 +\left|{\beta^4}_{16} \right|^2 ,\notag \\
    a^4_{1,10}
    &= {\beta^4}_1 {\beta^4}_9^* +{\beta^4}_2 {\beta^4}_{10}^* = a4_{10,1}^*, \notag \\
    a^4_{3,12}
    &= {\beta^4}_3 {\beta^4}_{11}^* +{\beta^4}_4 {\beta^4}_{12}^* = a4_{12,3}^* ,\notag\\
    a^4_{5,14}
    &= {\beta^4}_5 {\beta^4}_{13}^* +{\beta^4}_6 {\beta^4}_{14}^* = a4_{14,5}^*, \notag \\
    a^4_{7,16}
    &= {\beta^4}_7 {\beta^4}_{15}^* +{\beta^4}_8 {\beta^4}_{16}^* = a4_{16,7}^* . \label{a^4_ij}
\end{align}
From (\ref{B1}), by tracing out $\mathbf{JR}$, the reduced density matrix $\rho^4_{\mathbf{BH}}$ at step 4 is given by
\begin{align}
    \rho^4_{\mathbf{BH}} &=
    \operatorname{Tr}_{\mathbf{JR}}\!\left[ \rho^4_{\mathbf{BH}\cup \mathbf{JR}} \right] \notag \\
    &= \left[
\begin{array}{*{8}{c}}
a^4_{1,1} & 0 & 0 & 0 & 0 & 0 & 0 & 0  \\
0 & a^4_{3,3} & 0 & 0 & 0 & 0 & 0 & 0  \\
0 & 0 & a^4_{5,5} & 0 & 0 & 0 & 0 & 0  \\
0 & 0 & 0 & a^4_{7,7} & 0 & 0 & 0 & 0  \\
0 & 0 & 0 & 0 & a^4_{10,10} & 0 & 0 & 0  \\
0 & 0 & 0 & 0 & 0 & a^4_{12,12} & 0 & 0  \\
0 & 0 & 0 & 0 & 0 & 0 & a^4_{14,14} & 0  \\
0 & 0 & 0 & 0 & 0 & 0 & 0 & a^4_{16,16}
\end{array}
    \right].
\end{align}
Similarly, by tracing out $\mathbf{BH}$ in (\ref{B1}), we can obtain the reduced density matrix $\rho^4_{\mathbf{JR}}$ at step 4 to be
\begin{equation}
\resizebox{9cm}{!}{$
\begin{split}
    \rho^4_{\mathbf{JR}} &=
    \operatorname{Tr}_{\mathbf{BH}}\!\left[ \rho^4_{\mathbf{BH}\cup \mathbf{JR}} \right] \\
    &= \left[
\begin{array}{*{2}{c}}
a^4_{1,1} +a^4_{3,3} +a^4_{5,5} +a^4_{7,7} & 0 \\
0 & a^4_{10,10} +a^4_{12,12} +a^4_{14,14} +a^4_{16,16}
\end{array}
    \right].
\end{split}
$}
\end{equation}

On the other hand, from (\ref{B1}), by performing a partial transpose on $\mathbf{JR}$ and tracing out $\mathbf{ER}$, the partially transposed density matrix is given by
\begin{equation}
\resizebox{9cm}{!}{$
\begin{split}
    &\rho^{4,T_{\mathbf{JR}}}_{\mathbf{BH},\mathbf{JR}} = \operatorname{Tr}^{4,T_{\mathbf{JR}}}_{\mathbf{ER}}\!\left[ \ket{\psi^4}\bra{\psi^4} \right]
    = \\
    &\left[
\begin{array}{*{16}{c}}
a^4_{1,1} & 0 & 0 & 0 & 0 & 0 & 0 & 0 & 0 & 0 & 0 & 0 & 0 & 0 & 0 & 0 \\
0 & 0 & 0 & 0 & 0 & 0 & 0 & 0 & a^4_{1,10} & 0 & 0 & 0 & 0 & 0 & 0 & 0 \\
0 & 0 & a^4_{3,3} & 0 & 0 & 0 & 0 & 0 & 0 & 0 & 0 & 0 & 0 & 0 & 0 & 0 \\
0 & 0 & 0 & 0 & 0 & 0 & 0 & 0 & 0 & 0 & a^4_{3,12} & 0 & 0 & 0 & 0 & 0 \\
0 & 0 & 0 & 0 & a^4_{5,5} & 0 & 0 & 0 & 0 & 0 & 0 & 0 & 0 & 0 & 0 & 0 \\
0 & 0 & 0 & 0 & 0 & 0 & 0 & 0 & 0 & 0 & 0 & 0 & a^4_{5,14} & 0 & 0 & 0 \\
0 & 0 & 0 & 0 & 0 & 0 & a^4_{7,7} & 0 & 0 & 0 & 0 & 0 & 0 & 0 & 0 & 0 \\
0 & 0 & 0 & 0 & 0 & 0 & 0 & 0 & 0 & 0 & 0 & 0 & 0 & 0 & a^4_{7,16} & 0 \\
0 & a^4_{10,1} & 0 & 0 & 0 & 0 & 0 & 0 & 0 & 0 & 0 & 0 & 0 & 0 & 0 & 0 \\
0 & 0 & 0 & 0 & 0 & 0 & 0 & 0 & 0 & a^4_{10,10} & 0 & 0 & 0 & 0 & 0 & 0 \\
0 & 0 & 0 & a^4_{12,3} & 0 & 0 & 0 & 0 & 0 & 0 & 0 & 0 & 0 & 0 & 0 & 0 \\
0 & 0 & 0 & 0 & 0 & 0 & 0 & 0 & 0 & 0 & 0 & a^4_{12,12} & 0 & 0 & 0 & 0 \\
0 & 0 & 0 & 0 & 0 & a^4_{14,5} & 0 & 0 & 0 & 0 & 0 & 0 & 0 & 0 & 0 & 0 \\
0 & 0 & 0 & 0 & 0 & 0 & 0 & 0 & 0 & 0 & 0 & 0 & 0 & a^4_{14,14} & 0 & 0 \\
0 & 0 & 0 & 0 & 0 & 0 & 0 & a^4_{16,7} & 0 & 0 & 0 & 0 & 0 & 0 & 0 & 0 \\
0 & 0 & 0 & 0 & 0 & 0 & 0 & 0 & 0 & 0 & 0 & 0 & 0 & 0 & 0 & a^4_{16,16}
\end{array}
    \right],
\end{split}
$}
\end{equation}
where the matrix elements are shown in (\ref{a^4_ij}).

\subsection{At step 5}
\label{AppB_5}
From (\ref{psi5}), the density matrix $\rho^5_{\mathbf{BH}\cup \mathbf{JR}}$ is
\begin{align}
    \rho^5_{\mathbf{BH}\cup \mathbf{JR}} &= \operatorname{Tr}_{\mathbf{ER}}\!\left[ \ket{\psi^5}\bra{\psi^5} \right] \notag \\
    &= \left[
\begin{array}{*{16}{c}}
a^5_{1,1} & 0 & 0 & 0 & 0 & a^5_{1,6} & 0 & 0  \\
0 & a^5_{2,2} & 0 & 0 & a^5_{2,5} & 0 & 0 & 0  \\
0 & 0 & a^5_{3,3} & 0 & 0 & 0 & 0 & a^5_{3,8}  \\
0 & 0 & 0 & a^5_{4,4} & 0 & 0 & a^5_{4,7} & 0  \\
0 & a^5_{5,2} & 0 & 0 & a^5_{5,5} & 0 & 0 & 0  \\
a^5_{6,1} & 0 & 0 & 0 & 0 & a^5_{6,6} & 0 & 0  \\
0 & 0 & 0 & a^5_{7,4} & 0 & 0 & a^5_{7,7} & 0  \\
0 & 0 & a^5_{8,3} & 0 & 0 & 0 & 0 & a^5_{8,8}
\end{array}
    \right],
\label{B2}
\end{align}
where
\begin{align}
    a^5_{1,1} &= \sum_{n=1}^{4} \left|{\beta^5}_n \right|^2, \qquad\qquad
    a^5_{2,2} = \sum_{n=5}^{8} \left|{\beta^5}_n \right|^2, \notag \\
    a^5_{3,3} &= \sum_{n=9}^{12} \left|{\beta^5}_n \right|^2, \qquad\qquad
    a^5_{4,4} = \sum_{n=13}^{16} \left|{\beta^5}_n \right|^2, \notag \\
    a^5_{5,5} &= \sum_{n=17}^{20} \left|{\beta^5}_n \right|^2, \qquad\qquad
    a^5_{6,6} = \sum_{n=21}^{24} \left|{\beta^5}_n \right|^2, \notag \\
    a^5_{7,7} &= \sum_{n=25}^{28} \left|{\beta^5}_n \right|^2, \qquad\qquad
    a^5_{8,8} = \sum_{n=29}^{32} \left|{\beta^5}_n \right|^2, \notag \\
    a^5_{1,6} &= \sum_{n=1}^{4} {\beta^5}_n \left({\beta^5}_{20+n}\right)^* = {a^6_{6,1}}^*, \notag \\
    a^5_{2,5} &= \sum_{n=5}^{8} {\beta^5}_n \left({\beta^5}_{12+n}\right)^* = {a^6_{5,2}}^*, \notag \\
    a^5_{3,8} &= \sum_{n=9}^{12} {\beta^5}_n \left({\beta^5}_{20+n}\right)^* = {a^6_{8,3}}^*, \notag \\
    a^5_{4,7} &= \sum_{n=13}^{16} {\beta^5}_n \left({\beta^5}_{12+n}\right)^* = {a^6_{7,4}}^* \label{a^5_ij}
\end{align}
with ${\beta^5}_i$ in Table \ref{Tb2}.
The reduced density matrix $\rho^5_{\mathbf{BH}}$ is
\begin{align}
    \rho^5_{\mathbf{BH}} &= \operatorname{Tr}_{\mathbf{JR}}\!\left[ \rho^5_{\mathbf{BH}\cup \mathbf{JR}} \right] \notag \\
    &= \left[
\begin{array}{*{8}{c}}
a^5_{1,1} +a^5_{2,2} & 0 & 0 & 0  \\
0 & a^5_{3,3} +a^5_{4,4} & 0 & 0  \\
0 & 0 & a^5_{5,5} +a^5_{6,6} & 0  \\
0 & 0 & 0 & a^5_{7,7} +a^5_{8,8}
\end{array}
    \right].
\end{align}
%
%
%
The reduced density matrix $\rho^5_{\mathbf{JR}}$ is
\begin{equation}
\resizebox{8.5cm}{!}{$
\begin{split}
    \rho^5_{\mathbf{JR}} &= \operatorname{Tr}_{\mathbf{BH}}\!\left[ \rho^5_{\mathbf{BH}\cup \mathbf{JR}} \right] \\
    &= \left[
\begin{array}{*{2}{c}}
a^5_{1,1} +a^5_{3,3} +a^5_{5,5} +a^5_{7,7} & 0 \\
0 & a^5_{2,2} +a^5_{4,4} +a^5_{6,6} +a^5_{8,8}
\end{array}
    \right].
\end{split}
$}
\end{equation}
%
The partially transposed matrix is shown  to be
\begin{align}
    \rho^{5,T_{\mathbf{JR}}}_{\mathbf{BH},\mathbf{JR}} &= \operatorname{Tr}^{5,T_{\mathbf{JR}}}_{\mathbf{ER}}\!\left[ \ket{\psi^5}\bra{\psi^5} \right] \notag\\
    &= \left[
\begin{array}{*{16}{c}}
a^5_{1,1} & 0 & 0 & 0 & 0 & a^5_{2,5} & 0 & 0  \\
0 & a^5_{2,2} & 0 & 0 & a^5_{1,6} & 0 & 0 & 0  \\
0 & 0 & a^5_{3,3} & 0 & 0 & 0 & 0 & a^5_{4,7}  \\
0 & 0 & 0 & a^5_{4,4} & 0 & 0 & a^5_{3,8} & 0  \\
0 & a^5_{6,1} & 0 & 0 & a^5_{5,5} & 0 & 0 & 0  \\
a^5_{5,2} & 0 & 0 & 0 & 0 & a^5_{6,6} & 0 & 0  \\
0 & 0 & 0 & a^5_{8,3} & 0 & 0 & a^5_{7,7} & 0  \\
0 & 0 & a^5_{7,4} & 0 & 0 & 0 & 0 & a^5_{8,8}
\end{array}
    \right]
\end{align}
with the matrix elements shown in (\ref{a^5_ij}).

\subsection{At step 6}
\label{AppB_6}
From (\ref{psi6}), the density matrix $\rho^6_{\mathbf{BH}\cup \mathbf{JR}}$ is
\begin{align}
    \rho^6_{\mathbf{BH}\cup \mathbf{JR}} &= \operatorname{Tr}_{\mathbf{ER}}\!\left[ \ket{\psi^6}\bra{\psi^6} \right]
    = \left[
\begin{array}{*{8}{c}}
a^6_{1,1} & 0 & 0 & a^6_{1,4}  \\
0 & a^6_{2,2} & a^6_{2,3} & 0  \\
0 & a^6_{3,2} & a^6_{3,3} & 0  \\
a^6_{4,1} & 0 & 0 & a^6_{4,4}
\end{array}
    \right],
\label{B3}
\end{align}
where
\begin{align}
    a^6_{1,1}
    &= \sum_{n=1}^{16} \left|{\beta^6}_n \right|^2, \qquad\qquad
    a^6_{2,2}
    = \sum_{n=17}^{32} \left|{\beta^6}_n \right|^2, \notag \\
    a^6_{3,3}
    &= \sum_{n=33}^{48} \left|{\beta^6}_n \right|^2, \qquad\qquad
    a^6_{4,4}
    = \sum_{n=49}^{64} \left|{\beta^6}_n \right|^2, \notag \\
    a^6_{1,4}
    &= \sum_{n=1}^{16} {\beta^6}_n \left({\beta^6}_{48+n}\right)^* = {a^6_{4,1}}^*, \notag \\
    a^6_{2,3}
    &= \sum_{n=17}^{32} {\beta^6}_n \left({\beta^6}_{16+n}\right)^* = {a^6_{3,2}}^* \label{a^6_ij}
\end{align}
with the matrix elements dependent on ${\beta^6}_i$ in Table \ref{Tb3}.
The reduced density matrix $\rho^6_{\mathbf{BH}}$ is
\begin{align}
    \rho^6_{\mathbf{BH}} &= \operatorname{Tr}_{\mathbf{JR}}\!\left[ \rho^6_{\mathbf{BH}\cup \mathbf{JR}} \right]
    = \left[
\begin{array}{*{2}{c}}
a^6_{1,1} +a^6_{2,2} & 0 \\
0 & a^6_{3,3} +a^6_{4,4}
\end{array}
    \right].
\end{align}
%
%
%
The reduced density matrix $\rho^6_{\mathbf{JR}}$ is
\begin{align}
    \rho^6_{\mathbf{JR}} &= \operatorname{Tr}_{\mathbf{BH}}\!\left[ \rho^6_{\mathbf{BH}\cup \mathbf{JR}} \right]
    = \left[
\begin{array}{*{2}{c}}
a^6_{1,1} +a^6_{3,3} & 0 \\
0 & a^6_{2,2} +a^6_{4,4}
\end{array}
    \right].
\end{align}
%
The partially transposed matrix is shown as
\begin{align}
    \rho^{6,T_{\mathbf{JR}}}_{\mathbf{BH},\mathbf{JR}} &= \operatorname{Tr}^{6,T_{\mathbf{JR}}}_{\mathbf{ER}}\!\left[ \ket{\psi^6}\bra{\psi^6} \right]
    = \left[
\begin{array}{*{8}{c}}
a^6_{1,1} & 0 & 0 & a^6_{2,3}  \\
0 & a^6_{2,2} & a^6_{1,4} & 0  \\
0 & a^6_{4,1} & a^6_{3,3} & 0  \\
a^6_{3,2} & 0 & 0 & a^6_{4,4}
\end{array}
    \right] ,\label{rho6T}
\end{align}
where the matrix elements are shown in (\ref{a^6_ij}).

\subsection{At step 7}
\label{AppB_7}
From (\ref{psi7}), the density matrix $\rho^7_{\mathbf{BH}\cup \mathbf{JR}}$ is
\begin{align}
    \rho^7_{\mathbf{BH}\cup \mathbf{JR}} &= \operatorname{Tr}_{\mathbf{ER}}\!\left[ \ket{\psi^7}\bra{\psi^7} \right]
    = \left[
\begin{array}{*{2}{c}}
a^7_{1,1} & 0 \\
0 & a^7_{2,2}
\end{array}
    \right],
\end{align}
where
\begin{align}
    a^7_{1,1} &= \sum_{n=1}^{32} \left|{\beta^7}_n \right|^2, \qquad\qquad
    a^7_{2,2} = \sum_{n=33}^{64} \left|{\beta^7}_n \right|^2 . \label{a^7_ij}
\end{align}
The reduced density matrix $\rho^7_{\mathbf{BH}}$ at step 7 is
\begin{align}
    \rho^7_{\mathbf{BH}} &= \operatorname{Tr}_{\mathbf{JR}}\!\left[ \rho^7_{\mathbf{BH}\cup \mathbf{JR}} \right] = 0,
\end{align}
and $S_{\mathbf{BH}} =0$, since $\mathbf{BH}$ does not entangle with $\mathbf{JR}$ and $\mathbf{ER}$.
The reduced density matrix $\rho^7_{\mathbf{JR}}$ at step 7 is
\begin{align}
    \rho^7_{\mathbf{JR}} &= \operatorname{Tr}_{\mathbf{BH}}\!\left[ \rho^7_{\mathbf{BH}\cup \mathbf{JR}} \right]
    = \rho^7_{\mathbf{BH}\cup \mathbf{JR}} ,
\end{align}
%
and the partially-transposed matrix is also given by
\begin{align}
    \rho^{7,T_{\mathbf{JR}}}_{\mathbf{BH},\mathbf{JR}} &= \operatorname{Tr}^{7,T_{\mathbf{JR}}}_{\mathbf{ER}}\!\left[ \ket{\psi^7}\bra{\psi^7} \right]
    =\rho^7_{\mathbf{BH}\cup \mathbf{JR}} ,
\end{align}
which makes the mutual information and negativity vanish at the final step.



\end{document}